\documentstyle[thmsa,a4,sw20lart]{article}

\input{tcilatex}
\begin{document}

\title{{\sc Mechanics of Continuous Media in}{\bf \ }$(\overline{L}_n,g)${\sc %
-Spaces.}\\
{\bf \ III. Relative Accelerations }}
\author{S. Manoff \\
{\it Bulgarian Academy of Sciences,}\\
{\it \ Institute for Nuclear Research and Nuclear Energy,}\\
{\it \ Blvd. Tzarigradsko Chaussee 72,}\\
{\it \ 1784 Sofia - Bulgaria}}
\date{{\it E-mail address: smanov@inrne.bas.bg}}
\maketitle

\begin{abstract}
Basic notions of continuous media mechanics are introduced for spaces with
affine connections and metrics. The physical interpretation of the notion of
relative acceleration is discussed. The notions of deformation acceleration
, shear acceleration, rotation (vortex) acceleration, and expansion
acceleration are introduced. Their corresponding notions, generated by the
torsion and curvature, are considered. A classification is proposed for
auto-parallel vector fields with different kinematic characteristics.
Relations between the kinematic characteristics of the relative acceleration
and these of the relative velocity are found. A summary of the introduced
and considered notions is given. A classification is proposed related to the
kinematic characteristics of the relative velocity and the kinematic
characteristics related to the relative acceleration.

PACS numbers: 11.10.-z; 11.10.Ef; 7.10.+g; 47.75.+f; 47.90.+a; 83.10.Bb
\end{abstract}

\tableofcontents

\section{Introduction}

The notion of relative acceleration is very important for the modern
gravitational theories in (pseudo) Riemannian spaces or in spaces with one
affine connection and metrics. The theoretical basis for construction of
gravitational wave detectors is related to deviation equations describing
relative accelerations induced by the curvature. For continuous media
mechanics in $(\overline{L}_n,g)$-spaces the notion of relative acceleration
and its kinematic characteristics (deformation acceleration, shear
acceleration, rotation (vortex) acceleration, and expansion acceleration)
could be related to the change of the volume force in a continuous media.

In Section 2 the physical interpretation of the notion of relative
acceleration is discussed. The notions of deformation acceleration , shear
acceleration, rotation (vortex) acceleration, and expansion acceleration are
introduced. Their corresponding notions, generated by the torsion and
curvature, are considered. In Section 3 a classification is proposed for
auto-parallel vector fields with different kinematic characteristics. In
Section 4 relations between the kinematic characteristics of the relative
acceleration and these of the relative velocity are found. In Section 5 a
summary of the introduced and considered notions is given. In Section 6 a
classification is proposed related to the kinematic characteristics of the
relative velocity and the kinematic characteristics related to the relative
acceleration.

{\it Remark. }The present paper is the third part of a larger research
report on the subject with the title ''Contribution to continuous media
mechanics in $(\overline{L}_n,g)$-spaces'' with the following contents:

I. Introduction and mathematical tools.

II. Relative velocity and deformations.

III. Relative accelerations.

IV. Stress (tension) tensor.

The parts are logically self-dependent considerations of the main topics
considered in the report.

\section{Relative acceleration. Deformation acceleration, shear
acceleration, rotation (vortex) acceleration, and expansion acceleration}

\subsection{Physical interpretation of the notion of relative acceleration}

Let us consider the change of the vector $\nabla _u\xi _{(a)\perp }$

(a) along a curve $x^i(\tau ,\lambda _0^a=\,$const.$)$ and

(b) along a curve $x^i(\tau _0=\,$const.,\thinspace \thinspace \thinspace
\thinspace $\lambda ^a)$.

\subsubsection{Relative acceleration}

(a) In the first case 
\begin{eqnarray*}
\frac D{d\tau }(\frac{D\xi _{(a)\perp }}{d\tau })_{(\tau _0,\lambda _0^a)}
&=&(\nabla _u(\nabla _u\xi _{(a)\perp }))_{(\tau _0,\lambda _0^a)}=(\nabla
_u\nabla _u\xi _{(a)\perp })_{(\tau _0,\lambda _0^a)}= \\
&=&\stackunder{d\tau \rightarrow 0}{\lim }\frac{\nabla _u\xi _{(a)\perp
(\tau _0+d\tau ,\lambda _0^a)}-\nabla _u\xi _{(a)\perp (\tau _0,\lambda
_0^a)}}{d\tau }=
\end{eqnarray*}
\begin{equation}
=\stackunder{d\tau \rightarrow 0}{\lim }\frac{\left( \frac{D\xi _{(a)\perp }%
}{d\tau }\right) _{(\tau _0+d\tau ,\lambda _0^a)}-\left( \frac{D\xi
_{(a)\perp }}{d\tau }\right) _{(\tau _0,\lambda _0^a)}}{d\tau }\text{
\thinspace \thinspace .}  \label{7.1}
\end{equation}

The vector $\nabla _u\nabla _u\xi _{(a)\perp }$ has two components with
respect to the vector $u$: one component collinear to $u$ and one component
orthogonal to $u$, i.e. 
\begin{eqnarray}
\nabla _u\nabla _u\xi _{(a)\perp } &=&\frac 1e\cdot g(\nabla _u\nabla _u\xi
_{(a)\perp },u)\cdot u+\overline{g}[h_u(\nabla _u\nabla _u\xi _{(a)\perp })]=
\nonumber \\
&=&\frac{\overline{\overline{l}}_{(a)}}e\cdot u+\,_{rel}a_{(a)}\text{
\thinspace \thinspace \thinspace \thinspace ,}  \label{7.2}
\end{eqnarray}

\noindent where 
\begin{eqnarray*}
\overline{\overline{l}}_{(a)} &=&g(\nabla _u\nabla _u\xi _{(a)\perp },u)%
\text{ \thinspace \thinspace \thinspace \thinspace ,\thinspace \thinspace
\thinspace \thinspace \thinspace \thinspace \thinspace \thinspace }%
_{rel}a_{(a)}=\overline{g}[h_u(\nabla _u\nabla _u\xi _{(a)\perp })]\,\,\,\,\,%
\text{,} \\
\overline{\overline{l}}_{(a)} &=&g\left( \frac{D^2\xi _{(a)\perp }}{d\tau ^2}%
,u\right) \,\,\,\,\,\,\,\text{, \thinspace \thinspace \thinspace \thinspace
\thinspace \thinspace \thinspace }_{rel}a_{(a)}=\overline{g}[h_u\left( \frac{%
D^2\xi _{(a)\perp }}{d\tau ^2}\right) ]\text{ \thinspace \thinspace
\thinspace ,} \\
g(u,\,_{rel}a) &=&0\text{ \thinspace \thinspace .}
\end{eqnarray*}

The vector $\frac{D^2\xi _{(a)\perp }}{d\tau ^2}$ determines the change of
the velocity $\frac{D\xi _{(a)\perp }}{d\tau }$ at the point $(\tau
_0,\lambda _0^a)$ and, therefore, it describes the total acceleration of a
material point at the point $(\tau _0,\lambda _0^a)$%
\begin{eqnarray}
_{total}a_{(a)(\tau _0,\lambda _0^a)} &=&a_{(a)\parallel (\tau _0,\lambda
_0^a)}+a_{(a)\perp (\tau _0,\lambda _0^a)}\text{ \thinspace \thinspace
\thinspace ,}  \label{7.3} \\
a_{(a)\parallel (\tau _0,\lambda _0^a)} &=&\left( \frac{\overline{\overline{l%
}}}e\cdot u\right) _{(\tau _0,\lambda _0^a)}\,\text{\thinspace \thinspace
\thinspace \thinspace \thinspace \thinspace ,\thinspace \thinspace
\thinspace }  \nonumber \\
\text{\thinspace \thinspace \thinspace \thinspace }a_{(a)\perp (\tau
_0,\lambda _0^a)} &=&\,_{rel}a_{(a)(\tau _0,\lambda _0^a)}=\overline{g}[%
h_u\left( \frac{D^2\xi _{(a)\perp }}{d\tau ^2}\right) ]_{(\tau _0,\lambda
_0^a)}\,\,\,\,\text{.}  \nonumber
\end{eqnarray}

The total acceleration $_{total}a_{(a)}$ has two parts. The first part $%
a_{(a)\parallel }$ is the change of the velocity $D\xi _{(a)\perp }/d\tau $
along the flow. The second part $a_{(a)\perp }$ is the acceleration of the
material points orthogonal to the flow. The last acceleration is exactly the
relative acceleration $_{rel}a_{(a)}$ between material points lying at the
cross-section of the flow orthogonal to it. This is so, because $a_{(a)\perp
}$ for $d\tau \rightarrow 0$ is lying on the cross-section, orthogonal to $u$%
, where this cross-section is determined by the infinitesimal vectors $%
\overline{\xi }_{(a)\perp }=d\lambda ^a\cdot \xi _{(a)\perp }$, $a=1,\ldots
,n-1$, connecting the material point with co-ordinates $x^i(\tau _0,\lambda
_0^a)$ with the material points with co-ordinates $\{x^i(\tau _0,\lambda
_0^a+d\lambda ^a)\}$, $a=1,\ldots ,n-1$. So, the orthogonal to $u$ part $%
_{rel}a_{(a)}$ of the total acceleration $_{total}a_{(a)}$, written as $%
_{rel}a_{(a)}=a_{(a)\perp }$, 
\begin{equation}
\text{\thinspace }_{rel}a_{(a)}=\overline{g}[h_u\left( \frac{D^2\xi
_{(a)\perp }}{d\tau ^2}\right) ]\,\,\,\,\,\text{,}  \label{7.4}
\end{equation}

\noindent  has well grounds for its interpretation as relative acceleration
between material points of a flow, lying on a hypersurface orthogonal to the
velocity $u$ of the flow.

The representation of the relative acceleration by means of its different
parts, induced by the absolute acceleration, by the torsion, and by the
curvature leads to the corresponding interpretation of the different types
of relative accelerations in a flow. The different types of relative
accelerations could be expressed by the different types of velocities,
generated by the relative velocity. So we have a full picture of the forms
of relative accelerations and of the forms of relative velocities as well as
the relations between relative accelerations and relative velocities in a
continuous media.

\subsection{Kinematic characteristics connected with the notion relative
acceleration}

\subsubsection{Relative acceleration and its representation}

We will recall now some results already found in the considerations of the
relative acceleration and its kinematic characteristics from a more general
point of view \cite{Manoff-9}.

The notion {\it relative acceleration} vector field (relative acceleration) $%
_{rel}a$ can be defined (in analogous way as $_{rel}v$) as the orthogonal to
a non-isotropic vector field $u$ [$g(u,u)=e\neq 0$] projection of the second
covariant derivative (along the same non-isotropic vector field $u$) of
(another) vector field $\xi $, i.e. 
\begin{equation}
_{rel}a=\overline{g}(h_u(\nabla _u\nabla _u\xi ))=g^{ij}\cdot h_{\overline{j}%
\overline{k}}\cdot (\xi ^k\text{ }_{;l}\cdot u^l)_{;m}\cdot u^m\cdot e_i%
\text{ .}  \label{7.5}
\end{equation}

$\nabla _u\nabla _u\xi $ $=(\xi ^i$ $_{;l}\cdot u^l)_{;m}\cdot u^m\cdot e_i$
is the second covariant derivative of a vector field $\xi $ along the vector
field $u$. It is an essential part of all types of deviation equations in $%
V_n$-, $(L_n,g)$-, and $(\overline{L}_n,g)$-spaces \cite{Manoff-12} $\div $ 
\cite{Iliev-8}.

If we take into account the expression for $\nabla _u\xi $%
\[
\nabla _u\xi =k[g(\xi )]-\pounds _\xi u\text{,} 
\]

\noindent and differentiate covariant along $u$, then we obtain 
\[
\nabla _u\nabla _u\xi =\{\nabla _u[(k)g]\}(\xi )+(k)(g)(\nabla _u\xi
)-\nabla _u(\pounds _\xi u) 
\]

By means of the relations 
\[
k(g)\overline{g}=k\text{ ,} 
\]
\[
\nabla _u[k(g)]=(\nabla _uk)(g)+k(\nabla _ug)\text{ ,} 
\]
\[
\{\nabla _u[k(g)]\}\overline{g}=\nabla _uk+k(\nabla _ug)\overline{g}\text{ ,}
\]

\noindent $\nabla _u\nabla _u\xi $ can be written in the form 
\begin{equation}
\nabla _u\nabla _u\xi =\frac le\cdot H(u)+B(h_u)\xi -k(g)\pounds _\xi
u-\nabla _u(\pounds _\xi u)  \label{7.6}
\end{equation}

[compare with $\nabla _u\xi =\frac le\cdot a+k(h_u)\xi -\pounds _\xi u$],

\noindent where 
\[
H=B(g)=(\nabla _uk)(g)+k(\nabla _ug)+k(g)k(g)\text{ ,} 
\]
\[
B=\nabla _uk+k(g)k+k(\nabla _ug)\overline{g}=\nabla _uk+k(g)k-k(g)(\nabla _u%
\overline{g})\text{ .} 
\]

The orthogonal to $u$ covariant projection of $\nabla _u\nabla _u\xi $ will
have therefore the form 
\begin{equation}
h_u(\nabla _u\nabla _u\xi )=h_u[\frac le\cdot H(u)-k(g)\pounds _\xi u-\nabla
_u\pounds _\xi u]+[h_u(B)h_u](\xi )\text{ .}  \label{7.7}
\end{equation}

In the special case, when $g(u,\xi )=l=0$ and $\pounds _\xi u=0$ , the above
expression has the simple form 
\begin{equation}
h_u(\nabla _u\nabla _u\xi )=[h_u(B)h_u](\xi )=A(\xi )\text{ ,}  \label{7.8}
\end{equation}

[compare with $h_u(\nabla _u\xi )=[h_u(k)h_u](\xi )=d(\xi )$].

The explicit form of $H(u)$ follows from the explicit form of $H$ and its
action on the vector field $u$%
\begin{equation}
H(u)=(\nabla _uk)[g(u)]+k(\nabla _ug)(u)+k(g)(a)=\nabla _u[k(g)(u)]=\nabla
_ua\text{ .}  \label{7.9}
\end{equation}

Now $h_u[\nabla _u\nabla _u\xi ]$ can be written in the form 
\begin{equation}
h_u(\nabla _u\nabla _u\xi )=h_u[\frac le\cdot \nabla _ua-k(g)(\pounds _\xi
u)-\nabla _u(\pounds _\xi u)]+A(\xi )  \label{7.10}
\end{equation}

[compare with $h_u(\nabla _u\xi )=h_u(\frac le\cdot a-\pounds _\xi u)+d(\xi
) $].

The explicit form of $A=h_u(B)h_u$ can be found in analogous way as the
explicit form for $d=h_u(k)h_u$ in the expression for $_{rel}v$.

On the other side, by the use of the relations 
\begin{eqnarray}
\nabla _u(h_u(\nabla _u\xi )) &=&(\nabla _uh_u)(\nabla _u\xi )+h_u(\nabla
_u\nabla _u\xi )\text{ \thinspace \thinspace \thinspace ,}  \nonumber \\
h_u(\nabla _u\xi ) &=&g(_{rel}v)\text{ \thinspace \thinspace \thinspace ,} 
\nonumber \\
h_u(\nabla _u\nabla _u\xi ) &=&\nabla _u[g(_{rel}v)]-(\nabla _uh_u)(\nabla
_u\xi )=  \nonumber \\
&=&(\nabla _ug)(_{rel}v)+g(\nabla _u(_{rel}v))-(\nabla _uh_u)(\frac{%
\overline{l}}e\cdot u+\,_{rel}v)=  \nonumber \\
&=&(\nabla _ug)(_{rel}v)+g(\nabla _u(_{rel}v))-\frac{\overline{l}}e\cdot
(\nabla _uh_u)(u)-(\nabla _uh_u)(_{rel}v)\text{ \thinspace \thinspace
\thinspace ,}  \label{7.10a} \\
\nabla _uh_u &=&\nabla _ug-[u(\frac 1e)]\cdot g(u)\otimes g(u)-\frac 1e\cdot
[g(a)\otimes g(u)+g(u)\otimes g(a)+  \nonumber \\
&&+(\nabla _ug)(u)\otimes g(u)+g(u)\otimes (\nabla _ug)(u)\,\text{\thinspace
\thinspace \thinspace \thinspace ,}  \nonumber \\
(\nabla _uh_u)(_{rel}v) &=&(\nabla _ug)(_{rel}v)-\frac 1e\cdot
[g(a,\,_{rel}v)\cdot g(u)+(\nabla _ug)(u,\,_{rel}v)\cdot g(u)]\text{
\thinspace \thinspace \thinspace \thinspace \thinspace ,}  \nonumber \\
\nabla _u[g(u,\,_{rel}v)] &=&u[g(u,\,_{rel}v)]=0=(\nabla
_ug)(u,\,_{rel}v)+g(a,\,_{rel}v)+g(u,\nabla _u(_{rel}v))\text{ \thinspace ,}
\nonumber \\
(\nabla _uh_u)(_{rel}v) &=&(\nabla _ug)(_{rel}v)+\frac 1e\cdot g(u,\nabla
_u(_{rel}v))\cdot g(u)\text{ ,}  \nonumber
\end{eqnarray}
\begin{eqnarray}
(\nabla _uh_u)(u) &=&(\nabla _ug)(u)-e\cdot [u(\frac 1e)]\cdot g(u)- 
\nonumber \\
&&-\frac 1e\cdot [e\cdot g(a)+g(a,u)\cdot g(u)+e\cdot (\nabla
_ug)(u)+(\nabla _ug)(u,u)\cdot g(u)]\text{ ,}  \label{7.10b}
\end{eqnarray}

\noindent we can find $h_u(\nabla _u\nabla _u\xi )$ in the forms 
\begin{eqnarray}
h_u(\nabla _u\nabla _u\xi ) &=&h_u(\nabla _u(_{rel}v))-\frac 1e\cdot
[ul-g(a,\xi )-(\nabla _ug)(u,\xi )]\cdot  \nonumber \\
&&\cdot \{[\frac{ue}e-\frac 1e\cdot g(u,a)-\frac 1e\cdot (\nabla
_ug)(u,u)]\cdot g(u)-g(a)\}\text{\thinspace \thinspace \thinspace \thinspace
\thinspace \thinspace ,}  \label{7.10c}
\end{eqnarray}
\begin{eqnarray}
\overline{g}[h_u(\nabla _u\nabla _u\xi )] &=&\overline{g}[h_u(\nabla
_u(_{rel}v))]-\frac 1e\cdot [ul-g(a,\xi )-(\nabla _ug)(u,\xi )]\cdot 
\nonumber \\
&&\cdot \{[\frac{ue}e-\frac 1e\cdot g(u,a)-\frac 1e\cdot (\nabla
_ug)(u,u)]\cdot u-a\}\,\,\,\,\,\,\text{.}  \label{7.10d}
\end{eqnarray}

{\it Special case:} $\overline{V}_n$-spaces: $\nabla _ug=0$, $l=0$, $e=\,$%
const., $g(u,a)=0$. 
\begin{equation}
h_u(\nabla _u\nabla _u\xi )=h_u(\nabla _u(_{rel}v))-\frac 1e\cdot g(a,\xi
)\cdot g(a)\text{ \thinspace \thinspace \thinspace \thinspace \thinspace
\thinspace \thinspace .}  \label{7.10e}
\end{equation}

Under the additional condition $g(a,\xi ):=0$: $h_u(\nabla _u\nabla _u\xi
)=h_u(\nabla _u(_{rel}v))$.

Under the additional condition $a:=0$ (auto-parallel vector field $u$,
inertial flow): $h_u(\nabla _u\nabla _u\xi )=h_u(\nabla _u(_{rel}v))$.

\subsubsection{Deformation acceleration, shear acceleration, rotation
acceleration, and expansion acceleration}

The covariant tensor of second rank $A$, named {\it deformation acceleration}
tensor can be represented as a sum, containing three terms: a trace-free
symmetric term, an antisymmetric term and a trace term 
\begin{equation}
A=\,_sD+W+\frac 1{n-1}\cdot U\cdot h_u  \label{7.11}
\end{equation}

\noindent where 
\begin{equation}
D=h_u(_sB)h_u  \label{7.12}
\end{equation}
\begin{equation}
W=h_u(_aB)h_u  \label{7.13}
\end{equation}
\begin{equation}
U=\overline{g}[_sA]=\overline{g}[D]  \label{7.14}
\end{equation}
\begin{equation}
_sB=\frac 12\cdot (B^{ij}+B^{ji})\cdot e_i.e_j\text{ , }_aB=\frac 12\cdot
(B^{ij}-B^{ji})\cdot e_i\wedge e_j\text{,}  \label{7.15}
\end{equation}
\begin{equation}
_sA=\frac 12\cdot (A_{ij}+A_{ji})\cdot e^i.e^j\text{ },  \label{7.16}
\end{equation}
\begin{equation}
_sD=D-\frac 1{n-1}\cdot \overline{g}[D]\cdot h_u=D-\frac 1{n-1}\cdot U\cdot
h_u\text{ .}  \label{7.17}
\end{equation}

The shear-free symmetric tensor $_sD$ is the {\it shear acceleration} tensor
(shear acceleration), the antisymmetric tensor $W$ is the {\it rotation
acceleration} tensor (rotation acceleration) and the invariant $U$ is the 
{\it expansion acceleration }invariant (expansion acceleration).
Furthermore, every one of these quantities can be divided into three parts:
torsion- and curvature-free acceleration, acceleration induced by torsion
and acceleration induced by curvature.

Let us now consider the representation of every acceleration quantity in its
essential parts, connected with its physical interpretation.

The deformation acceleration tensor $A$ can be written in the following
forms 
\begin{equation}
A=\,_sD+W+\frac 1{n-1}\cdot U\cdot h_u=A_0+G=\,_FA_0-\,_TA_0+G\text{ ,}
\label{7.18}
\end{equation}
\begin{equation}
A=\,_sD_0+W_0+\frac 1{n-1}\cdot U_0\cdot h_u+\,_sM+N+\frac 1{n-1}\cdot
I\cdot h_u\text{ ,}  \label{7.19}
\end{equation}
\begin{equation}
\begin{array}{c}
A=\,_{sF}D_0+_FW_0+\frac 1{n-1}\cdot _FU_0\cdot h_u- \\ 
-(_{sT}D_0+\,_TW_0+\frac 1{n-1}\cdot _TU\cdot h_u)+ \\ 
+\,_sM+N+\frac 1{n-1}\cdot I\cdot h_u\text{ ,}
\end{array}
\label{7.20}
\end{equation}

\noindent where 
\begin{equation}
A_0=\,_FA_0-\,_TA_0=\,_sD_0+W_0+\frac 1{n-1}\cdot U_0\cdot h_u\text{ ,}
\label{7.21}
\end{equation}
\begin{equation}
_FA_0=\,_{sF}D_0+\,_FW_0+\frac 1{n-1}\cdot \,_FU_0\cdot h_u\text{ ,}
\label{7.22}
\end{equation}
\begin{equation}
_FA_0(\xi )=h_u(\nabla _{\xi _{\perp }}a)\text{ ,\thinspace \thinspace
\thinspace \thinspace \thinspace \thinspace \thinspace \thinspace \thinspace
\thinspace \thinspace \thinspace \thinspace \thinspace \thinspace }\xi
_{\perp }=\overline{g}[h_u(\xi )]\text{ ,}  \label{7.23}
\end{equation}
\begin{equation}
_TA_0=\,_{sT}D_0+\,_TW_0+\frac 1{n-1}\cdot _TU_0\cdot h_u\text{ ,}
\label{7.24}
\end{equation}
\begin{equation}
G=\,_sM+N+\frac 1{n-1}\cdot I\cdot h_u=h_u(K)h_u\text{ ,}  \label{7.25}
\end{equation}
\begin{equation}
h_u([R(u,\xi )]u)=h_u(K)h_u(\xi )\,\,\,\,\,\,\,\,\,\,\,\,\,\,\,\,\,\,\,\text{
for }\forall \text{ }\xi \in T(M)\text{ , }  \label{7.26}
\end{equation}
\begin{equation}
\lbrack R(u,\xi )]u=\nabla _u\nabla _\xi u-\nabla _\xi \nabla _uu-\nabla
_{\pounds _u\xi }u\text{ ,}  \label{7.27}
\end{equation}
\begin{equation}
K=K^{kl}\cdot e_k\otimes e_l\text{ ,\thinspace \thinspace \thinspace
\thinspace \thinspace \thinspace \thinspace \thinspace \thinspace \thinspace
\thinspace \thinspace }K^{kl}=R^k\text{ }_{mnr}\cdot g^{rl}\cdot u^m\cdot u^n%
\text{ ,}  \label{7.28}
\end{equation}

$R^k$ $_{mnr}$ are the components of the contravariant Riemannian curvature
tensor, 
\begin{eqnarray}
K_a &=&K_a^{kl}\cdot e_k\wedge e_l\text{, \thinspace \thinspace \thinspace
\thinspace \thinspace \thinspace \thinspace \thinspace \thinspace \thinspace
\thinspace \thinspace \thinspace \thinspace \thinspace \thinspace \thinspace 
}K_a^{kl}=\frac 12\cdot (K^{kl}-K^{lk})\text{ , }  \label{7.29a} \\
K_s &=&K_s^{kl}\cdot e_k.e_l\text{ , \thinspace \thinspace \thinspace
\thinspace \thinspace \thinspace \thinspace \thinspace \thinspace \thinspace
\thinspace \thinspace \thinspace \thinspace \thinspace \thinspace \thinspace
\thinspace \thinspace \thinspace \thinspace \thinspace }K_s^{kl}=\frac
12\cdot (K^{kl}+K^{lk})\text{ ,}  \label{7.29b}
\end{eqnarray}
\begin{equation}
_sD=\,_sD_0+\,_sM\text{ , \thinspace \thinspace \thinspace \thinspace
\thinspace \thinspace \thinspace \thinspace \thinspace \thinspace \thinspace
\thinspace \thinspace \thinspace \thinspace \thinspace }W=W_0+N=\,_FW_0-%
\,_TW_0+N\text{ ,}  \label{7.30}
\end{equation}
\begin{equation}
U=U_0+I=\,_FU_0-\,_TU_0+I\text{ ,}  \label{7.31}
\end{equation}
\begin{equation}
_sM=M-\frac 1{n-1}\cdot I\cdot h_u\text{ ,\thinspace \thinspace \thinspace
\thinspace \thinspace \thinspace }M=h_u(K_s)h_u\text{ ,\thinspace \thinspace
\thinspace \thinspace }I=\overline{g}[M]=g^{\overline{i}\overline{j}}\cdot
M_{ij}\text{ ,}  \label{7.32}
\end{equation}
\begin{equation}
N=h_u(K_a)h_u\text{ ,}  \label{7.33}
\end{equation}
\begin{equation}
_sD_0=\,_{sF}D_0-\,_{sT}D_0=\,_FD_0-\frac 1{n-1}\cdot \,_FU_0\cdot
h_u-(_TD_0-\frac 1{n-1}\cdot \,_TU_0\cdot h_u)\text{ ,}  \label{7.34}
\end{equation}
\begin{equation}
_sD_0=\,_{sF}D_0-\,_TD_0-\frac 1{n-1}\cdot (_FU_0-\,_TU_0)\cdot h_u\text{ ,}
\label{7.35}
\end{equation}
\begin{equation}
_sD_0=D_0-\frac 1{n-1}\cdot U_0\cdot h_u\text{ ,}  \label{7.36}
\end{equation}
\begin{equation}
_{sF}D_0=\,_FD_0-\frac 1{n-1}\cdot \,_FU_0\cdot h_u\text{ , \thinspace
\thinspace \thinspace \thinspace \thinspace \thinspace \thinspace \thinspace
\thinspace \thinspace \thinspace \thinspace \thinspace \thinspace \thinspace
\thinspace }_FD_0=h_u(b_s)h_u\text{ ,}  \label{7.37}
\end{equation}
\begin{equation}
b=b_s+b_a\text{ ,\thinspace \thinspace \thinspace \thinspace \thinspace
\thinspace \thinspace \thinspace \thinspace }b=b^{kl}\cdot e_k\otimes e_l%
\text{ ,\thinspace \thinspace \thinspace \thinspace \thinspace \thinspace
\thinspace \thinspace \thinspace }b^{kl}=a^k\text{ }_{;n}\cdot g^{nl}\text{ ,%
}  \label{7.38}
\end{equation}
\begin{equation}
a^k=u^k\text{ }_{;m}\cdot u^m\text{ , \thinspace \thinspace \thinspace
\thinspace \thinspace \thinspace \thinspace \thinspace \thinspace }%
b_s=b_s^{kl}\cdot e_k.e_l\text{ , \thinspace \thinspace \thinspace
\thinspace \thinspace }b_s^{kl}=\frac 12\cdot (b^{kl}+b^{lk})\text{ ,}
\label{7.39}
\end{equation}
\begin{equation}
b_a=b_a^{kl}\cdot e_k\wedge e_l\text{ ,\thinspace \thinspace \thinspace
\thinspace \thinspace \thinspace \thinspace \thinspace \thinspace \thinspace
\thinspace \thinspace \thinspace \thinspace \thinspace \thinspace \thinspace
\thinspace \thinspace }b_a^{kl}=\frac 12\cdot (b^{kl}-b^{lk})\text{ ,}
\label{7.40}
\end{equation}
\begin{equation}
_FU_0=\overline{g}[_FD_0]=g[b]-\frac 1e\cdot g(u,\nabla _ua)\text{
,\thinspace \thinspace \thinspace \thinspace \thinspace \thinspace
\thinspace \thinspace \thinspace \thinspace }g[b]=g_{\overline{k}\overline{l}%
}\cdot b^{kl}\text{ ,}  \label{7.41}
\end{equation}
\begin{equation}
_{sT}D_0=\,_TD_0-\frac 1{n-1}\cdot \,_TU_0\cdot h_u=\,_{sF}D_0-\,_sD_0\text{
, \thinspace \thinspace \thinspace \thinspace \thinspace }_TD_0=\,_FD_0-D_0%
\text{ ,}  \label{7.42}
\end{equation}
\begin{equation}
U_0=\overline{g}[D_0]=\,_FU_0-\,_TU_0\text{ ,\thinspace \thinspace
\thinspace \thinspace \thinspace \thinspace \thinspace \thinspace \thinspace
\thinspace \thinspace \thinspace \thinspace \thinspace \thinspace \thinspace
\thinspace \thinspace \thinspace \thinspace \thinspace \thinspace \thinspace 
}_TU_0=\overline{g}[_TD_0]\text{ ,}  \label{7.43}
\end{equation}
\begin{equation}
_FW_0=h_u(b_a)h_u\text{ , \thinspace \thinspace \thinspace \thinspace
\thinspace \thinspace \thinspace \thinspace \thinspace \thinspace \thinspace
\thinspace \thinspace }_TW_0=\,_FW_0-W_0\text{ .}  \label{7.44}
\end{equation}

Under the conditions $\pounds _\xi u=0$ , $\xi =\xi _{\perp }=\overline{g}[%
h_u(\xi )]$ , ($l=0$), the expression for $h_u(\nabla _u\nabla _u\xi )$ can
be written in the forms 
\begin{equation}
h_u(\nabla _u\nabla _u\xi _{\perp })=A(\xi _{\perp })=A_0(\xi _{\perp
})+G(\xi _{\perp })\text{ ,}  \label{7.45}
\end{equation}
\begin{equation}
h_u(\nabla _u\nabla _u\xi _{\perp })=\,_FA_0(\xi _{\perp })-\,_TA_0(\xi
_{\perp })+G(\xi _{\perp })\text{ ,}  \label{7.46}
\end{equation}
\[
h_u(\nabla _u\nabla _u\xi _{\perp })=(_{sF}D_0+\,_FW_0+\frac 1{n-1}\cdot
\,_FU_0\cdot g)(\xi _{\perp })- 
\]
\begin{equation}
-(_{sT}D_0+\,_TW_0+\frac 1{n-1}\cdot \,_TU_0\cdot g)(\xi _{\perp
})+(_sM+N+\frac 1{n-1}\cdot I\cdot g)(\xi _{\perp })\text{ ,}  \label{7.47}
\end{equation}

\noindent which enable one to find a physical interpretation of the
quantities $_sD$,$W$,$U$ and of the contained in their structure quantities$%
_{sF}D_0$, $_FW_0$, $_FU_0$, $_{sT}D_0$, $_TW_0$, $_TU_0$, $_sM$, $N$, $I$.
The individual designation, connected with their physical interpretation, is
given in the Section 11 - Table $1$. The expressions of these quantities in
terms of the kinematic characteristics of the relative velocity are given in
a section below.

After the above consideration the following proposition can be formulated:

\begin{proposition}
The covariant vector{\bf \ }$g(_{rel}a)=h_u(\nabla _u\nabla _u\xi )$ can be
written in the form 
\[
g(_{rel}a)=h_u[\frac le\cdot \nabla _ua-\nabla _{\pounds _\xi u}u-\nabla
_u(\pounds _\xi u)+T(\pounds _\xi u,u)]+A(\xi )\text{ ,} 
\]
\end{proposition}

\noindent where 
\begin{equation}
A(\xi )=\,_sD(\xi )+W(\xi )+\frac 1{n-1}\cdot U\cdot h_u(\xi )\text{ .}
\label{7.48}
\end{equation}

For the case of affine symmetric connection [$T(w,v)=0$ for $\forall $ $%
w,v\in T(M)$ , $T_{ij}^k=0$, $\Gamma _{ij}^k=\Gamma _{ji}^k$ ] and
Riemannian metric [$\nabla _vg=0$ for $\forall v\in T(M)$, $g_{ij;k}=0$]
kinematic characteristics are obtained in $V_n$-spaces, connected with the
notion relative velocity \cite{Ehlers}, \cite{Stephani}, \cite{Manoff-14}
and relative acceleration \cite{Manoff-13}. For the case of affine
non-symmetric connection [$T(w,v)\neq 0$ for $\forall $ $w,v\in T(M)$ , $%
\Gamma _{jk}^i\neq \Gamma _{kj}^i$] and Riemannian metric kinematic
characteristics are obtained in $U_n$-spaces \cite{Manoff-13}.

\section{Classification of auto-parallel vector fields on the basis of the
kinematic characteristics connected with the relative velocity and relative
acceleration}

The classification of (pseudo) Riemannian spaces $V_n$, admitting the
existence of auto-parallel (in the case of $V_n$-spaces they are geodesic)
vector fields ($\nabla _uu=a=0$) with given kinematic characteristics,
connected with the notion relative velocity \cite{Ehlers}, can be extended
to a classification of differentiable manifolds with contravariant and
covariant affine connections and metrics, admitting auto-parallel vector
fields with certain kinematic characteristics, connected with the relative
velocity and the relative acceleration. In this way, the following two
schemes for the existence of special type $1$ and $2$ of vector fields can
be proposed (s. Table $2$). Different types of combinations between the
single conditions of the two schemes can also be taken under consideration.

\subsection{Inertial flows. Special geodesic vector fields with vanishing
kinematic characteristics, induced by the curvature, in (pseudo) Riemannian
spaces}

On the basis of the classification $2$ the following propositions in the
case of $V_n$-spaces can be proved:

\begin{proposition}
Non-isotropic geodesic vector fields in $V_n$-spaces are geodesic vector
fields with curvature rotation acceleration tensor $N$ equal to zero, i.e. $%
N=0$.
\end{proposition}

Proof: 
\begin{equation}
\begin{array}{c}
N=h_u(K_a)h_u=h_{ik}\cdot K_a^{kl}\cdot h_{lj}\cdot e^i\wedge e^j, \\ 
K_a^{kl}=\frac 12\cdot (K^{kl}-K^{lk})=\frac 12\cdot (R_{mnr}^k\cdot
g^{rl}-R_{mnr}^l\cdot g^{rk})\cdot u^m\cdot u^n,
\end{array}
\label{8.1}
\end{equation}

For the case of $V_n$-space, where 
\begin{equation}
R_{kmnr}=R_{nrkm}\text{ , }R_{kmnr}=g_{kl}\cdot R^l\text{ }_{mnr}\text{ ,}
\label{8.2}
\end{equation}

\noindent the conditions 
\begin{equation}
R^k\text{ }_{mnr}\cdot g^{rl}=R^l\text{ }_{nmr}\cdot g^{rk}\text{
,\thinspace \thinspace \thinspace \thinspace \thinspace \thinspace
\thinspace \thinspace \thinspace \thinspace \thinspace \thinspace \thinspace
\thinspace }R^k\text{ }_{mn}\text{ }^l=R^l\text{ }_{nm}\text{ }^k\text{,}
\label{8.3}
\end{equation}

\noindent follow and therefore

\begin{equation}
K_a^{kl}=\frac 12\cdot (R^k\text{ }_{mnr}\cdot g^{rl}-R^l\text{ }_{mnr}\cdot
g^{rk})\cdot u^m\cdot u^n=0\text{ , }  \label{8.4}
\end{equation}
\[
K_a=0\text{ , }N=0\text{ .} 
\]

\begin{proposition}
Non-isotropic geodesic vector fields in $V_n$-spaces with equal to zero
Ricci tensor ($R_{ik}=R^l$ $_{ikl}=g_m^l\cdot R^m$ $_{ikl}=0$) are geodesic
vector fields with curvature rotation acceleration $N$ and curvature
expansion acceleration $I$, both equal to zero, i.e. $N=0$, $I=0$.
\end{proposition}

Proof: $1$. From the above proposition, it follows that $K_a=0$ and $N=0$. 
\begin{eqnarray}
2\text{. }I &=&g[K]=g_{ij}\cdot K^{ij}=g_{ij}\cdot R^i\text{ }_{mnr}\cdot
g^{rj}\cdot u^m\cdot u^n=  \nonumber \\
&=&g_i^r\cdot R^i\text{ }_{mnr}\cdot u^m\cdot u^n=R_{mn}\cdot u^m\cdot u^n=0%
\text{ .}  \label{8.5}
\end{eqnarray}

\begin{proposition}
Non-isotropic geodesic vector fields in $V_n$-spaces with constant curvature 
\begin{equation}
[R(\xi ,\eta )]v=\frac 1{n\cdot (n-1)}\cdot R_0\cdot [g(v,\xi )\cdot \eta
-g(v,\eta )\cdot \xi ]\text{ , \thinspace \thinspace \thinspace \thinspace
\thinspace \thinspace \thinspace \thinspace }\forall \xi ,\eta ,v\in T(M)%
\text{,}  \label{8.6}
\end{equation}
\end{proposition}

[in index form 
\begin{equation}
R^i\text{ }_{jkl}=\frac{R_0}{n\cdot (n-1)}\cdot (g_l^i\cdot
g_{jk}-g_k^i\cdot g_{jl})\text{ , \thinspace \thinspace \thinspace
\thinspace \thinspace \thinspace \thinspace \thinspace \thinspace \thinspace
\thinspace \thinspace \thinspace }R_0=\text{const.]}  \label{8.7}
\end{equation}

{\it are geodesic vector fields with curvature shear acceleration and
curvature rotation acceleration, both equal to zero, i.e.} $_sM=0$, $N=0$.

Proof: $1$. From the first proposition in this subsection above, it follows
that $N=0$. 
\[
2\text{. }_sM=M-\frac 1{n-1}\cdot I\cdot h_u\text{ , \thinspace \thinspace
\thinspace \thinspace \thinspace \thinspace \thinspace \thinspace \thinspace
\thinspace \thinspace \thinspace \thinspace \thinspace \thinspace \thinspace 
}M=h_u(K_s)h_u=g(K_s)g\text{ ,} 
\]
\[
M=g(K_s)g=g_{ik}\cdot K^{kl}\cdot g_{lj}\cdot e^i.e^j=M_{ij}\cdot e^i.e^j%
\text{ ,} 
\]
\begin{equation}
M_{ij}=g_{ik}\cdot R^k\text{ }_{mnr}\cdot g^{rl}\cdot g_{lj}\cdot u^m\cdot
u^n=R_{imnj}\cdot u^m\cdot u^n=\frac{R_0}{n\cdot (n-1)}\cdot e\cdot h_{ij}%
\text{ ,}  \label{8.8}
\end{equation}
\begin{equation}
M=\frac{R_0\cdot e}{n\cdot (n-1)}\cdot h_u\text{ ,\thinspace \thinspace
\thinspace \thinspace \thinspace \thinspace \thinspace \thinspace \thinspace
\thinspace \thinspace \thinspace \thinspace \thinspace \thinspace \thinspace
\thinspace }e=g(u,u)=g_{ij}\cdot u^i\cdot u^j\text{ ,}  \label{8.9}
\end{equation}
\begin{equation}
I=g[K]=\overline{g}[M]=g^{ij}\cdot M_{ij}=\frac 1n\cdot R_0\cdot e\text{
,\thinspace \thinspace \thinspace \thinspace \thinspace \thinspace
\thinspace \thinspace \thinspace \thinspace \thinspace \thinspace \thinspace
\thinspace \thinspace \thinspace \thinspace \thinspace \thinspace }%
g^{ij}\cdot h_{ij}=n-1\text{ ,}  \label{8.10}
\end{equation}
\[
_sM=M-\frac 1{n-1}\cdot I\cdot h_u=0\text{.} 
\]

The projections of the curvature tensor of the type $G=h_u(K)h_u$ (or $R^i$ $%
_{jkl}\cdot u^j\cdot u^k$) along the non-isotropic vector field $u$ acquire
a natural physical meaning as quantities, connected with the kinematic
characteristics curvature shear acceleration $_sM$, curvature rotation
acceleration $N$, and curvature expansion acceleration $I$.

The projection of the Ricci tensor ($g[K]$, or $R_{ik}\cdot u^i\cdot u^k$)
and the Raychaudhuri identity for vector fields represent an expression of
the curvature expansion acceleration, given in terms of the kinematic
characteristics of the relative velocity 
\begin{equation}
\begin{array}{c}
I=\overline{g}[M]=R_{ij}\cdot u^i\cdot u^j= \\ 
=-a^j\text{ }_{;j}+g^{\overline{i}\overline{j}}\cdot \,_sE_{ik}\cdot g^{%
\overline{k}\overline{l}}\cdot \sigma _{lj}+g^{\overline{i}\overline{j}%
}\cdot S_{ik}\cdot g^{\overline{k}\overline{l}}\cdot \omega _{lj}+\theta
_0^{.}+\frac 1{n-1}\cdot \theta _0\cdot \theta + \\ 
+\frac 1e\cdot [a^k\cdot (e_{,k}-u_{\overline{n}}\cdot T_{km}^n\cdot
u^m-g_{mn;k}\cdot u^{\overline{m}}\cdot u^{\overline{n}}-g_{\overline{k}%
\overline{m};l}\cdot u^l\cdot u^m)+ \\ 
+\frac 12\cdot (u^k\cdot e_{,k})_{;l}\cdot u^l-\frac 12\cdot (g_{mn;k}\cdot
u^k)_{;l}\cdot u^l\cdot u^{\overline{m}}\cdot u^{\overline{n}}]- \\ 
\frac 1{e^2}\cdot [\frac 34\cdot (e_{,k}\cdot u^k)^2-(e_{,k}\cdot u^k)\cdot
g_{mn;l}\cdot u^l\cdot u^{\overline{m}}\cdot u^{\overline{n}}+\frac 14\cdot
(g_{mn;l}\cdot u^l\cdot u^{\overline{m}}\cdot u^{\overline{n}})^2]\text{ ,}
\\ 
\theta ^{.}=\theta _{,k}\cdot u^k\,\,\,\,\text{.}
\end{array}
\label{8.12}
\end{equation}

In the case of $V_n$-spaces the kinematic characteristics, connected with
the relative velocity and the relative acceleration have the forms:

a) kinematic characteristics, connected with the relative velocity

\begin{center}
$
\begin{array}{ccc}
d=d_0 & d_1=0 & k=k_o \\ 
\sigma =\,_sE & _sP=0 & m=0 \\ 
\omega =S & Q=0 & q=0 \\ 
\theta =\theta _o & \theta _1=0 & \nabla _uu=a\neq 0\text{ , }a=0
\end{array}
$
\end{center}

b) kinematic characteristics of a non-inertial flow, connected with the
relative acceleration ($\nabla _uu=a\neq 0$)

\begin{center}
$
\begin{array}{ccc}
A=\,_FA_0+G & _TA_0=0 & N=0 \\ 
G=\,_sM+\frac 1{n-1}\cdot I\cdot h_u & _{sT}D_0=0 &  \\ 
W=\,_FW_0 & _TW_0=0 &  \\ 
U=\,_FU_0+I & _TU_0=0 & \nabla _uu=a\neq 0
\end{array}
$
\end{center}

c) kinematic characteristics of an inertial flow, connected with the
relative acceleration ($\nabla _uu=a=0$)

\begin{center}
$
\begin{array}{c}
A=G \\ 
G=\,_sM+\frac 1{n-1}\cdot I\cdot h_u \\ 
W=0 \\ 
U=I
\end{array}
\begin{array}{cc}
_TA_0=0 & N=0 \\ 
_{sT}D_0=0 &  \\ 
_TW_0=0 &  \\ 
_TU_0=0 & \nabla _uu=a=0
\end{array}
$
\end{center}

On the basis of the different kinematic characteristics dynamic systems and
flows can be classified and considered in $V_n$-spaces.

\subsection{Special flows with tangent vector fields over $(\overline{L}%
_n,g) $-spaces with vanishing kinematic characteristics induced by the
curvature}

The explicit forms of the quantities $G$, $M$, $N$ and $I$, connected with
accelerations induced by curvature, can be used for finding conditions for
existence of special types of contravariant vector fields with vanishing
characteristics induced by the curvature. $G$, $M$, $N$ and $I$ can be
expressed in the following forms: 
\begin{equation}
\begin{array}{c}
G=h_u(K)h_u=g(K)g-\frac 1e\cdot g(u)\otimes [g(u)](K)g\text{ , } \\ 
K[g(u)]=0\text{ ,}
\end{array}
\label{8.13}
\end{equation}
\begin{equation}
\begin{array}{c}
M=h_u(K_s)h_u=g(K_s)g-\frac 1{2e}\cdot \{g(u)\otimes
[g(u)](K)g+[g(u)](K)g\otimes g(u)\}= \\ 
=M_{ij}\cdot dx^i.dx^j=M_{\alpha \beta }\cdot e^\alpha .e^\beta \text{ ,
\thinspace \thinspace \thinspace \thinspace \thinspace \thinspace \thinspace
\thinspace \thinspace \thinspace \thinspace \thinspace \thinspace \thinspace
\thinspace \thinspace \thinspace \thinspace \thinspace \thinspace \thinspace
\thinspace \thinspace }M_{ij}=M_{ji}\text{ ,} \\ 
M_{ij}=\frac 12\cdot [g_{i\overline{k}}\cdot g_{\overline{l}j}+g_{j\overline{%
k}}\cdot g_{\overline{l}i}-\frac 1e\cdot (u_i\cdot g_{\overline{l}%
j}+u_j\cdot g_{\overline{l}i})\cdot u_{\overline{k}}]\cdot R^k\text{ }%
_{mnq}\cdot u^m\cdot u^n\cdot g^{ql}\text{ ,}
\end{array}
\label{8.14}
\end{equation}
\begin{equation}
I=\overline{g}[M]=g[K_s]=g[K]=R_{\rho \sigma }\cdot u^\rho \cdot u^\sigma
=R_{kl}\cdot u^k\cdot u^l\text{ ,}  \label{8.15}
\end{equation}
\begin{equation}
\begin{array}{c}
N=h_u(K_a)h_u=g(K_a)g-\frac 1{2e}\cdot \{g(u)\otimes
[g(u)](K)g-[g(u)](K)g\otimes g(u)\}= \\ 
=N_{ij}\cdot dx^i\wedge dx^j=N_{\alpha \beta }\cdot e^\alpha \wedge e^\beta 
\text{ ,\thinspace \thinspace \thinspace \thinspace \thinspace \thinspace
\thinspace \thinspace \thinspace \thinspace \thinspace \thinspace \thinspace
\thinspace \thinspace \thinspace \thinspace \thinspace \thinspace \thinspace
\thinspace \thinspace \thinspace \thinspace \thinspace \thinspace \thinspace 
}N_{ij}=-N_{ji}\text{ ,} \\ 
N_{ij}=\frac 12\cdot [g_{i\overline{k}}\cdot g_{\overline{l}j}-g_{j\overline{%
k}}\cdot g_{\overline{l}i}-\frac 1e\cdot (u_i\cdot g_{\overline{l}%
j}-u_j\cdot g_{\overline{l}i})\cdot u_{\overline{k}}]\cdot R^k\text{ }%
_{mnq}\cdot u^m\cdot u^n\cdot g^{ql}\text{ .}
\end{array}
\label{8.16}
\end{equation}

By means of the above expressions conditions can be found under which some
of the quantities $M$, $N$, $I$ vanish.

\subsubsection{Flows with tangent vector fields without rotation
acceleration, induced by the curvature ($N=0$)}

If the rotation acceleration $N$, induced by the curvature vanishes, i.e. if 
$N=0$, then the following proposition can be proved:

\begin{proposition}
The necessary and sufficient condition for the existence of a flow with
tangent vector field $u$ [$g(u,u)=e\neq 0$] without rotation acceleration,
induced by the curvature (i.e. with $N=0$), is the condition 
\begin{equation}
K_a=\frac 1{2e}\cdot \{u\otimes [g(u)](K)-[g(u)](K)\otimes u\}\text{ .}
\label{8.17}
\end{equation}
\end{proposition}

Proof: 1. Sufficiency: From the above expression it follows 
\[
\begin{array}{c}
N=h_u(K_a)h_u= \\ 
=g(K_a)g-\frac 1{2e}\cdot \{g(u)\otimes [g(u)](K)g-[g(u)](K)g\otimes g(u)\}=0%
\text{ ,} \\ 
g([g(u)](K))=[g(u)](K)g\text{ .}
\end{array}
\]

2. Necessity: If $N=h_u(K_a)h_u=0$, then 
\[
\begin{array}{c}
g(K_a)g=\frac 1{2e}\cdot \{g(u)\otimes [g(u)](K)g-[g(u)](K)g\otimes g(u)\}%
\text{ ,} \\ 
K_a=\frac 1{2e}\cdot \{u\otimes [g(u)](K)-[g(u)](K)\otimes u\}\text{ .}
\end{array}
\]

In co-ordinate basis the necessary and sufficient condition has the forms 
\begin{equation}
\begin{array}{c}
K^{ij}=K^{ji}+\frac 1e\cdot u_{\overline{l}}\cdot (u^i\cdot K^{lj}-u^j\cdot
K^{li})\text{ ,} \\ 
\{R_{\overline{j}nim}-R_{\overline{i}mjn}-\frac 1e\cdot (u_{\overline{i}%
}\cdot R_{\overline{l}mnj}-u_{\overline{j}}\cdot R_{\overline{l}mni})\cdot
u^l\}\cdot u^m\cdot u^n=0\text{ , }
\end{array}
\label{8.18}
\end{equation}

\noindent where 
\[
R_{\overline{i}jkl}=g_{\overline{i}\overline{n}}\cdot R^n\text{ }_{jkl}\text{
.} 
\]

\begin{proposition}
A sufficient condition for the existence of a flow with tangent vector field 
$u$ [$g(u,u)=e\neq 0$] without rotation acceleration, induced by the
curvature (i.e. with $N=0$), is the condition 
\begin{equation}
K_a=0\text{ .}  \label{8.19}
\end{equation}
\end{proposition}

Proof: From $K_a=0$ and the form for $N$, $N=h_u(K_a)h_u$, it follows $N=0$.

In co-ordinate basis 
\begin{equation}
\begin{array}{c}
(R^i\text{ }_{klm}\cdot g^{mj}-R^j\text{ }_{klm}\cdot g^{mi})\cdot u^k\cdot
u^l=0\text{ ,} \\ 
(R_{\overline{i}kjl}-R_{\overline{j}lik})\cdot u^k\cdot u^l=0\text{ .}
\end{array}
\label{8.20}
\end{equation}

$K_a=0$ can be presented also in the form 
\[
\lbrack g(\xi )]([R(u,v)]u)-[g(v)]([R(u,\xi )]u)=0\text{ ,\thinspace
\thinspace \thinspace \thinspace \thinspace \thinspace \thinspace \thinspace
\thinspace \thinspace \thinspace \thinspace \thinspace \thinspace \thinspace
\thinspace \thinspace \thinspace \thinspace }\forall \xi ,v\in T(M)\text{ .} 
\]

In this case $M=G=g(K)g$ , $I=\overline{g}[G]$.

\begin{proposition}
A sufficient condition for the existence of a flow with tangent vector field 
$u$ [$g(u,u)=e\neq 0$] without rotation acceleration, induced by the
curvature (i.e. with $N=0$), is the condition 
\begin{equation}
g(\eta ,[R(\xi ,v)]w)=g(\xi ,[R(\eta ,w)]v)\text{ , \thinspace \thinspace
\thinspace \thinspace \thinspace \thinspace \thinspace \thinspace \thinspace
\thinspace \thinspace \thinspace \thinspace \thinspace \thinspace \thinspace
\thinspace \thinspace \thinspace \thinspace \thinspace \thinspace \thinspace
\thinspace \thinspace \thinspace }\forall \eta ,\xi ,v,w\in T(M)\text{ ,}
\label{8.21}
\end{equation}
\end{proposition}

\noindent or in co-ordinate basis 
\begin{equation}
R_{\overline{i}jkl}=R_{\overline{k}lij}\text{ .}  \label{8.22}
\end{equation}

Proof: Because of $R(\xi ,u)=-R(u,\xi )$ and for $\eta =v$ the last
expression will be identical with the sufficient condition from the above
proposition.

\begin{remark}
If the rotation velocity $\omega $ vanishes $(\omega =0)$ for an
auto-parallel $(\nabla _uu=0)$ contravariant non-null vector field $u$, then
the rotation acceleration tensor $W$ will have the form \cite{Manoff-9} 
\[
W=\frac 12\cdot [h_u(\nabla _u\overline{g})\sigma -\sigma (\nabla _u%
\overline{g})h_u]\text{ .} 
\]
\end{remark}

From the last expression it is obvious that under the above conditions the
nonmetricity $(\nabla _ug\neq 0)$ in a $(\overline{L}_n,g)$-space is
responsible for nonvanishing the rotation acceleration $W$.

\subsubsection{Contravariant vector fields without shear acceleration $_sM$,
induced by the curvature ($_sM=0$)}

\begin{proposition}
The necessary and sufficient condition for the existence of a contravariant
vector field $u$ [$g(u,u)=e\neq 0$] without shear acceleration, induced by
the curvature (i.e. with $_sM=0$), is the condition 
\begin{equation}
M=\frac 1{n-1}\cdot I\cdot h_u=\frac 1{n-1}\cdot \overline{g}[M]\cdot h_u%
\text{ .}  \label{8.23}
\end{equation}
\end{proposition}

Proof: 1. Sufficiency: From the expression for $M$ and the definition of $%
_sM=M-\frac 1{n-1}\cdot I\cdot h_u$ it follows $_sM=0$.

2. Necessity: From $_sM=0=M-\frac 1{n-1}\cdot I\cdot h_u$ the form of $M$
follows.

In co-ordinate basis the necessary and sufficient condition can be written
in the form 
\begin{equation}
\begin{array}{c}
\{[g_{i\overline{k}}\cdot g_{\overline{l}j}+g_{j\overline{k}}\cdot g_{%
\overline{l}i}-\frac 1e\cdot (u_i\cdot g_{\overline{l}j}+u_j\cdot g_{%
\overline{l}i})\cdot u_{\overline{k}}]\cdot R^k\text{ }_{mns}\cdot g^{sl}-
\\ 
-\frac 2{n-1}\cdot R_{mn}\cdot (g_{ij}-\frac 1e\cdot u_i\cdot u_j)\}\cdot
u^m\cdot u^n=0\text{ .}
\end{array}
\label{8.24}
\end{equation}

The condition $_sM=0$ is identical with the condition for $K_s$: 
\begin{equation}
K_s=\frac 1{n-1}\cdot I\cdot h^u+\frac 1{2\cdot e}\cdot \{u\otimes
[g(u)](K)+[g(u)](K)\otimes u\}\text{ .}  \label{8.25}
\end{equation}

\subsubsection{Contravariant vector fields without shear and expansion
acceleration, induced by the curvature ($_sM=0$, $I=0$)}

\begin{proposition}
A sufficient condition for the existence of a contravariant vector field $u$
[$g(u,u)=e\neq 0$] without shear and expansion acceleration, induced by the
curvature (i.e. with $_sM=0$, $I=0$), is the condition 
\begin{equation}
K_s=\frac 1{2\cdot e}\cdot \{u\otimes [g(u)](K)+[g(u)](K)\otimes u\}\text{ .}
\label{8.26}
\end{equation}
\end{proposition}

Proof: After acting on the left and on the right side of the last expression
with $g$%
\[
\begin{array}{c}
g(K_s)g=\frac 1{2\cdot e}\cdot \{g(u)\otimes [g(u)](K)g+[g(u)](K)g\otimes
g(u)\}\text{ ,} \\ 
g([g(u)](K))=([g(u)](K))g=[g(u)](K)g\text{ ,\thinspace \thinspace \thinspace
\thinspace \thinspace \thinspace \thinspace \thinspace \thinspace \thinspace
\thinspace \thinspace }u(g)=g(u)\text{ ,}
\end{array}
\]
and comparing the result with the form for $M$, 
\[
M=h_u(K_s)h_u=g(K_s)g-\frac 1{2\cdot e}\cdot \{g(u)\otimes
[g(u)](K)g+[g(u)](K)g\otimes g(u)\}\text{ ,} 
\]
it follows that $M=0$. Since $I=\overline{g}[M]$, it follows that $I=0$ and $%
_sM=0$.

\begin{proposition}
A sufficient condition for the existence of a contravariant vector field $u$
[$g(u,u)=e\neq 0$] without shear and expansion acceleration, induced by the
curvature (i.e. with $_sM=0$, $I=0$), is the condition 
\[
K_s=0\text{ .} 
\]
\end{proposition}

Proof: From the condition and the form of $M$, $M=h_u(K_s)h_u$, it follows
that $M=0$ and therefore $I=0$ and $_sM=0$.

\subsubsection{Contravariant vector fields without shear and rotation
acceleration, induced by the curvature ($_sM=0$, $N=0$)}

\begin{proposition}
A sufficient condition for the existence of a contravariant vector field $u$
[$g(u,u)=e\neq 0$] without shear and rotation acceleration, induced by the
curvature (i.e. with $_sM=0$ , $N=0$), is the condition 
\begin{equation}
\begin{array}{c}
[R(u,\xi )]v=\frac R{n\cdot (n-1)}\cdot [g(v,u)\cdot \xi -g(v,\xi )\cdot u]%
\text{ , } \\ 
\forall v,\xi \in T(M)\text{ , \thinspace \thinspace \thinspace \thinspace
\thinspace \thinspace \thinspace \thinspace \thinspace \thinspace \thinspace
\thinspace \thinspace \thinspace \thinspace \thinspace \thinspace \thinspace 
}R\in C^r(M)\text{ .}
\end{array}
\label{8.27}
\end{equation}
\end{proposition}

Proof: Since $v$ is an arbitrary contravariant vector field it can be chosen
as $u$. Then, because of the relation 
\begin{equation}
h_u([R(u,\xi )]u)=h_u(K)h_u(\xi )=G(\xi )\text{ ,}  \label{8.28}
\end{equation}

\noindent it follows that 
\begin{equation}
G=h_u(K)h_u=\frac R{n\cdot (n-1)}\cdot e\cdot h_u=G_s\text{ ,\thinspace
\thinspace \thinspace \thinspace \thinspace \thinspace \thinspace \thinspace 
}G_a=h_u(K_a)h_u=0\text{ .}  \label{8.29}
\end{equation}

Therefore 
\begin{eqnarray}
M &=&G_s=\frac{R\cdot e}{n\cdot (n-1)}\cdot h_u\text{ , \thinspace
\thinspace \thinspace \thinspace \thinspace }N=G_a=0\text{ , }  \label{8.30a}
\\
I &=&\frac 1n\cdot R\cdot e\text{ , \thinspace \thinspace \thinspace
\thinspace \thinspace \thinspace \thinspace \thinspace \thinspace \thinspace
\thinspace \thinspace \thinspace \thinspace \thinspace \thinspace \thinspace
\thinspace \thinspace \thinspace \thinspace \thinspace \thinspace \thinspace
\thinspace \thinspace \thinspace \thinspace \thinspace \thinspace \thinspace
\thinspace \thinspace \thinspace \thinspace \thinspace \thinspace \thinspace
\thinspace \thinspace \thinspace \thinspace \thinspace \thinspace }_sM=0%
\text{ .}  \label{8.30b}
\end{eqnarray}

In co-ordinate basis the sufficient condition can be written in the form 
\begin{equation}
R^i\text{ }_{jkl}=\frac R{n\cdot (n-1)}\cdot (g_l^i\cdot g_{\overline{j}%
\overline{k}}-g_k^i\cdot g_{\overline{j}\overline{l}})  \label{8.31}
\end{equation}

\noindent and the following relations are fulfilled 
\begin{equation}
\begin{array}{c}
R_{jk}=R^l\text{ }_{jkl}=g_i^l\cdot R^i\text{ }_{jkl}=\frac 1n\cdot R\cdot
g_{\overline{j}\overline{k}}\text{ ,} \\ 
R=g^{jk}\cdot R_{jk}\text{ ,} \\ 
I=R_{jk}\cdot u^j\cdot u^k=\frac 1n\cdot R\cdot e\text{ .}
\end{array}
\label{8.32}
\end{equation}

\begin{proposition}
The necessary and sufficient conditions for the existence of $K$ in the form 
\begin{equation}
K=\frac 1{n-1}\cdot g[K]\cdot h^u  \label{8.33}
\end{equation}
\end{proposition}

\noindent are the conditions 
\[
_sM=0\text{ ,\thinspace \thinspace \thinspace \thinspace \thinspace
\thinspace \thinspace \thinspace \thinspace \thinspace \thinspace \thinspace
\thinspace \thinspace \thinspace \thinspace \thinspace }K_a=0\text{ .} 
\]

Proof: 1. Sufficiency: From $K_a=0$ it follows that $K=K_s$, $N=0$ and $%
M=g(K_s)g=g(K)g$. Therefore, $I=\overline{g}[M]=g[K]$. From $_sM=M-\frac
1{n-1}\cdot I\cdot h_u=0$, it follows that $M=\frac 1{n-1}\cdot g[K]\cdot
h_u=g(K)g$. From the last expression, it follows the above condition for $K$.

2. Necessity: From the condition $K=\frac 1{n-1}\cdot g[K]\cdot h^u$, it
follows that $K=K_s$ and therefore $K_a=0$, $N=0$ and $M=\frac 1{n-1}\cdot
g[K]\cdot h_u$, $I=g[K]$ [because of $h_u(h^u)h_u=h_u$, $h_u(\overline{g}%
)h_u=h_u$]. From the forms of $M$ and $I$, it follows that $_sM=0$.

\begin{proposition}
A sufficient condition for the existence of a contravariant vector field $u$
[$g(u,u)=e\neq 0$] without shear and rotation acceleration, induced by the
curvature (i.e. with $_sM=0$ , $N=0$), is the condition 
\[
K=\frac 1{n-1}\cdot g[K]\cdot h^u\text{ .} 
\]
\end{proposition}

Proof: Follows immediately from the above proposition.

\subsubsection{Contravariant vector fields without expansion acceleration,
induced by the curvature ($I=0$)}

By means of the covariant metric $g$ and the tensor field $K(v,\xi )$ the
notion contravariant Ricci tensor $Ricci$ can be introduced 
\begin{equation}
Ricci(v,\xi )=g[K(v,\xi )]\text{ ,\thinspace \thinspace \thinspace
\thinspace \thinspace \thinspace \thinspace \thinspace \thinspace \thinspace 
}\forall v,\xi \in T(M)\text{ ,}  \label{8.34}
\end{equation}

\noindent where 
\begin{equation}
K(v,\xi )=R^i\text{ }_{jkl}\cdot g^{lm}\cdot v^j\cdot \xi ^k\cdot \partial
_i\otimes \partial _m=R^\alpha \text{ }_{\beta \gamma \kappa }\cdot
g^{\kappa \delta }\cdot v^\beta \cdot \xi ^\gamma \cdot e_\alpha \otimes
e_\delta \text{ ,}  \label{8.35}
\end{equation}

\noindent and the following relations are fulfilled 
\begin{equation}
\begin{array}{c}
Ricci(e_\alpha ,e_\beta )=g[K(e_\alpha ,e_\beta )]=R_{\alpha \beta }\text{ ,}
\\ 
Ricci(\partial _i,\partial _j)=g[K(\partial _i,\partial _j)]=R_{ij}\text{ ,}
\\ 
Ricci(u,u)=g[K(u,u)]=g[K]=I\text{ .}
\end{array}
\label{8.36}
\end{equation}

\begin{proposition}
The necessary and sufficient condition for the existence of a contravariant
vector field $u$ [$g(u,u)=e\neq 0$] without expansion acceleration, induced
by the curvature (i.e. with $I=0$), is the condition 
\[
Ricci(u,u)=0\text{ .} 
\]
\end{proposition}

Proof: It follows immediately from the relation $Ricci(u,u)=g[K(u,u)]=g[K]=I$%
.

\begin{proposition}
A sufficient condition for the existence of a contravariant vector field $u$
[$g(u,u)=e\neq 0$] without expansion acceleration, induced by the curvature
(i.e. with $I=0$), is the condition 
\begin{equation}
\begin{array}{c}
Ricci(e_\alpha ,e_\beta )=R_{\alpha \beta }=R^\gamma \text{ }_{\alpha \beta
\gamma }=0\text{ ,} \\ 
Ricci(\partial _i,\partial _j)=R_{ij}=R^l\text{ }_{ijl}=0\text{ .}
\end{array}
\label{8.37}
\end{equation}
\end{proposition}

Proof: From $Ricci(\partial _i,\partial _j)=R_{ij}=0$, it follows that 
\[
R_{ij}\cdot u^i\cdot u^j=u^i\cdot u^j\cdot Ricci(\partial _i,\partial
_j)=Ricci(u,u)=I=0. 
\]

In a non-co-ordinate basis the proof is analogous to that in a co-ordinate
basis.

The existence of contravariant vector fields with vanishing characteristics,
induced by the curvature, is important for mathematical models of
gravitational interactions in theories over $(\overline{L}_n,g)$-spaces.

\section{Kinematic characteristics connected with the relative acceleration
and expressed in terms of the kinematic characteristics connected with the
relative velocity}

In this section the relations between the kinematic characteristics
connected with the relative velocity and the kinematic characteristics
connected with the relative acceleration are found. A summary of the
definitions of the kinematic characteristics is also given in Table 1..

The deformation, shear, rotation and expansion accelerations can be
expressed in terms of the shear, rotation and expansion velocity.

(a) Deformation acceleration tensor%
\index{deformation@deformation!deformation acceleration tensor@deformation
acceleration tensor} $A$: 
\begin{equation}
\begin{array}{c}
A=\frac 1e\cdot h_u(a)\otimes h_u(a)+\sigma (%
\overline{g})\sigma +\omega (\overline{g})\omega +\frac 2{n-1}\cdot \theta
\cdot (\sigma +\omega )+\frac 1{n-1}\cdot (\theta ^{.}+\frac{\theta ^2}{n-1}%
)\cdot h_u+ \\ 
\\ 
+\sigma (\overline{g})\omega +\omega (\overline{g})\sigma +\nabla _u\sigma
+\nabla _u\omega +\frac 1e\cdot h_u(a)\otimes (g(u))(2\cdot k-\nabla _u%
\overline{g})h_u+ \\ 
\\ 
+\frac 1e\cdot [\sigma (a)\otimes g(u)+g(u)\otimes \sigma (a)]+\frac 1e\cdot
[\omega (a)\otimes g(u)-g(u)\otimes \omega (a)]+ \\ 
\\ 
+h_u(\nabla _u\overline{g})\sigma +h_u(\nabla _u\overline{g})\omega \text{ ,}
\end{array}
\label{9.1}
\end{equation}

\noindent where 
\[
k=\varepsilon +s-(m+q)=k_0-(m+q)\text{ ,} 
\]
\begin{equation}
k(g)\pounds _\xi u=\nabla _{\pounds _\xi u}u-T(\pounds _\xi u,u)\text{ .}
\label{9.2}
\end{equation}

In index form 
\begin{equation}
\begin{array}{c}
A_{ij}=\frac 1e\cdot h_{i\overline{k}}\cdot a^k\cdot a^l\cdot h_{\overline{l}%
j}+\sigma _{ik}\cdot g^{\overline{k}\overline{l}}\cdot \sigma _{lj}+\omega
_{ik}\cdot g^{\overline{k}\overline{l}}\cdot \omega _{lj}+\frac 2{n-1}\cdot
\theta \cdot \sigma _{ij}+ \\ 
+\frac 1{n-1}\cdot (\theta ^{.}+\frac{\theta ^2}{n-1})\cdot h_{ij}+ \\ 
+\sigma _{ij;k}\cdot u^k+\frac 1e\cdot a^k\cdot [\sigma _{i\overline{k}%
}\cdot u_j+\sigma _{j\overline{k}}\cdot u_i+h_{\overline{k}(i}\cdot h_{j)%
\overline{l}}\cdot u_{\overline{n}}\cdot (2\cdot k^{nl}-g^{nl}\text{ }%
_{;r}\cdot u^r)]+ \\ 
+\frac 12\cdot (h_{i\overline{k}}\cdot g^{kl}\text{ }_{;r}\cdot u^r\cdot
\sigma _{\overline{l}j}+h_{j\overline{k}}\cdot g^{kl}\text{ }_{;r}\cdot
u^r\cdot \sigma _{\overline{l}i})+ \\ 
+\frac 12\cdot (h_{i\overline{k}}\cdot g^{kl}\text{ }_{;r}\cdot u^r\cdot
\omega _{\overline{l}j}+h_{j\overline{k}}\cdot g^{kl}\text{ }_{;r}\cdot
u^r\cdot \omega _{\overline{l}i})+ \\ 
+\sigma _{ik}\cdot g^{\overline{k}\overline{l}}\cdot \omega _{lj}-\sigma
_{jk}\cdot g^{\overline{k}\overline{l}}\cdot \omega _{li}+\frac 2{n-1}\cdot
\theta \cdot \omega _{ij}+\omega _{ij;r}\cdot u^r+ \\ 
+\frac 1e\cdot a^k\cdot [\omega _{i\overline{k}}\cdot u_j-\omega _{j%
\overline{k}}\cdot u_i+h_{\overline{k}[i}\cdot h_{j]\overline{l}}\cdot u_{%
\overline{n}}\cdot (2\cdot k^{nl}-g^{nl}\text{ }_{;r}\cdot u^r)]+ \\ 
+\frac 12\cdot (h_{i\overline{k}}\cdot g^{kl}\text{ }_{;r}\cdot u^r\cdot
\sigma _{\overline{l}j}-h_{j\overline{k}}\cdot g^{kl}\text{ }_{;r}\cdot
u^r\cdot \sigma _{\overline{l}i})+ \\ 
+\frac 12\cdot (h_{i\overline{k}}\cdot g^{kl}\text{ }_{;r}\cdot u^r\cdot
\omega _{\overline{l}j}-h_{j\overline{k}}\cdot g^{kl}\text{ }_{;r}\cdot
u^r\cdot \omega _{\overline{l}i})= \\ 
=D_{ij}+W_{ij}\text{ , \thinspace \thinspace \thinspace \thinspace }\theta
^{.}=u\theta =u^i\cdot \partial _i\theta =\theta _{,i}\cdot u^i\text{ ,}
\end{array}
\label{9.3}
\end{equation}
\[
A_{(ij)}=\frac 12\cdot (A_{ij}+A_{ji})\text{ , \thinspace \thinspace
\thinspace \thinspace \thinspace \thinspace \thinspace }A_{[ij]}=\frac
12\cdot (A_{ij}-A_{ji)}\text{ .} 
\]

(b) Shear acceleration tensor%
\index{shear@shear!shear acceleration tensor@shear acceleration tensor} $%
_sD=D-\frac 1{n-1}\cdot U\cdot h_u$: 
\begin{equation}
\begin{array}{c}
D=\frac 1e\cdot h_u(a)\otimes h_u(a)+\sigma (%
\overline{g})\sigma +\omega (\overline{g})\omega + \\ 
+\frac 2{n-1}\cdot \theta \cdot \sigma +\frac 1{n-1}\cdot (\theta ^{.}+\frac{%
\theta ^2}{n-1})\cdot h_u+\nabla _u\sigma + \\ 
+\frac 1{2\cdot e}\cdot [h_u(a)\otimes (g(u))(2\cdot k-\nabla _u\overline{g}%
)h_u+h_u((g(u))(2\cdot k-\nabla _u\overline{g}))\otimes h_u(a)]+ \\ 
+\frac 1e\cdot [\sigma (a)\otimes g(u)+g(u)\otimes \sigma (a))]+ \\ 
+\frac 12\cdot [h_u(\nabla _u\overline{g})\sigma +\sigma (\nabla _u\overline{%
g})h_u]+\frac 12\cdot [h_u(\nabla _u\overline{g})\omega -\omega (\nabla _u%
\overline{g})h_u]\text{ .}
\end{array}
\label{9.4}
\end{equation}

In index form 
\begin{equation}
\begin{array}{c}
D_{ij}=D_{ji}=\frac 1e\cdot h_{i\overline{k}}\cdot a^k\cdot a^l\cdot h_{%
\overline{l}j}+\sigma _{i\overline{k}}\cdot g^{kl}\cdot \sigma _{\overline{l}%
j}+\omega _{i\overline{k}}\cdot g^{kl}\cdot \omega _{\overline{l}j}+ \\ 
+\frac 2{n-1}\cdot \theta \cdot \sigma _{ij}+\frac 1{n-1}\cdot (\theta ^{.}+%
\frac{\theta ^2}{n-1})\cdot h_{ij}+\sigma _{ij;k}\cdot u^k+ \\ 
+\frac 1e\cdot a^k\cdot \{\sigma _{i\overline{k}}\cdot u_j+\sigma _{j%
\overline{k}}\cdot u_i+ \\ 
+h_{\overline{k}(i}\cdot h_{j)\overline{l}}\cdot [g^{ml}\cdot
(e_{,m}-g_{rs;m}\cdot u^{\overline{r}}\cdot u^{\overline{s}}-2\cdot
T_{mr}\,^n\cdot u^r\cdot u_{\overline{n}})-u_{\overline{n}}\cdot g^{nl}\text{
}_{;m}\cdot u^m]\}+ \\ 
+\frac 12\cdot (h_{ik}\cdot g^{\overline{k}\overline{l}}\text{ }_{;m}\cdot
u^m\cdot \sigma _{lj}+h_{jk}\cdot g^{\overline{k}\overline{l}}\text{ }%
_{;m}\cdot u^m\cdot \sigma _{li})+ \\ 
+\frac 12\cdot (h_{ik}\cdot g^{\overline{k}\overline{l}}\text{ }_{;m}\cdot
u^m\cdot \omega _{lj}+h_{jk}\cdot g^{\overline{k}\overline{l}}\text{ }%
_{;m}\cdot u^m\cdot \omega _{li})\text{ .}
\end{array}
\label{9.5}
\end{equation}

(c) Rotation acceleration tensor%
\index{rotation@rotation!rotation acceleration tensor@rotation acceleration
tensor} $W$%
\begin{equation}
\begin{array}{c}
W=\sigma (%
\overline{g})\omega +\omega (\overline{g})\sigma +\frac 2{n-1}\cdot \theta
\cdot \omega +\nabla _u\omega + \\ 
+\frac 1e\cdot [\omega (a)\otimes g(u)-g(u)\otimes \omega (a)]+ \\ 
+\frac 1{2\cdot e}\cdot [h_u(a)\otimes (g(u))(2\cdot k-\nabla _u\overline{g}%
)h_u-h_u((g(u))(2\cdot k-\nabla _u\overline{g}))\otimes h_u(a)]+ \\ 
+\frac 12\cdot [h_u(\nabla _u\overline{g})\sigma -\sigma (\nabla _u\overline{%
g})h_u]+\frac 12\cdot [h_u(\nabla _u\overline{g})\omega +\omega (\nabla _u%
\overline{g})h_u]\text{ .}
\end{array}
\label{9.6}
\end{equation}

In index form 
\begin{equation}
\begin{array}{c}
W_{ij}=-W_{ji}=\sigma _{ik}\cdot g^{\overline{k}\overline{l}}\cdot \omega
_{lj}-\sigma _{jk}\cdot g^{\overline{k}\overline{l}}\cdot \omega _{li}+\frac
2{n-1}\cdot \theta \cdot \omega _{ij}+\omega _{ij;k}\cdot u^k+ \\ 
+\frac 1e\cdot a^k\cdot \{\omega _{i\overline{k}}\cdot u_j-\omega _{j%
\overline{k}}\cdot u_i+h_{\overline{k}[i}\cdot h_{j]\overline{l}}\cdot
[g^{ml}\cdot (e_{,m}-g_{rs;m}\cdot u^{\overline{r}}\cdot u^{\overline{s}}-
\\ 
-2\cdot T_{mr}\,^n\cdot u^r\cdot u_{\overline{n}})-u_{\overline{n}}\cdot
g^{nl}\text{ }_{;m}\cdot u^m]\}+ \\ 
+\frac 12\cdot (h_{ik}\cdot g^{\overline{k}\overline{l}}\text{ }_{;m}\cdot
u^m\cdot \sigma _{lj}-h_{jk}\cdot g^{\overline{k}\overline{l}}\text{ }%
_{;m}\cdot u^m\cdot \sigma _{li})+ \\ 
+\frac 12\cdot (h_{i\overline{k}}\cdot g^{kl}\text{ }_{;m}\cdot u^m\cdot
\omega _{\overline{l}j}-h_{j\overline{k}}\cdot g^{kl}\text{ }_{;m}\cdot
u^m\cdot \omega _{\overline{l}i})\text{ .}
\end{array}
\label{9.7}
\end{equation}

(d) Expansion acceleration%
\index{expansion@expansion!expansion acceleration@expansion acceleration} $U$%
\begin{equation}
\begin{array}{c}
U=\frac 1e\cdot g(a,a)+%
\overline{g}[\sigma (\overline{g})\sigma ]+\overline{g}[\omega (\overline{g}%
)\omega ]+\theta ^{.}+\frac 1{n-1}\cdot \theta ^2+ \\ 
+\,\,\,\,\,\frac 1e\cdot [2\cdot g(u,\nabla _au)-2\cdot g(u,T(a,u))+(\nabla
_ug)(a,u)] \\ 
-\frac 1{e^2}\cdot g(u,a)\cdot [3\cdot g(u,a)+(\nabla _ug)(u,u)]\text{ .}
\end{array}
\label{9.8}
\end{equation}

In index form 
\begin{equation}
\begin{array}{c}
U=\frac 1e\cdot g_{\overline{i}\overline{j}}\cdot a^i\cdot a^j+g^{\overline{i%
}\overline{j}}\cdot g^{\overline{k}\overline{l}}\cdot \sigma _{ik}\cdot
\sigma _{jl}-g^{\overline{i}\overline{j}}\cdot g^{\overline{k}\overline{l}%
}\cdot \omega _{ik}\cdot \omega _{jl}+ \\ 
+\theta ^{.}+\frac 1{n-1}\cdot \theta ^2+ \\ 
+\frac 1e\cdot g_{\overline{k}\overline{l}}\cdot a^k\cdot [g^{ml}\cdot
(e_{,m}-g_{rs;m}\cdot u^{\overline{r}}\cdot u^{\overline{s}}-2\cdot
T_{mr}\,^n\cdot u^r\cdot u_{\overline{n}})- \\ 
-\,u_{\overline{n}}\cdot g^{nl}\text{ }_{;m}\cdot u^m]- \\ 
+\frac 1{e^2}\cdot [\frac 34\cdot (e_{,k}\cdot u^k)^2-(e_{,l}\cdot u^l)\cdot
g_{ij;k}\cdot u^k\cdot u^{\overline{i}}\cdot u^{\overline{j}}+ \\ 
+\,\frac 14\cdot (g_{ij;k}\cdot u^k\cdot u^{\overline{i}}\cdot u^{\overline{j%
}})^2]\text{ .}
\end{array}
\label{9.9}
\end{equation}

(e) Torsion-free and curvature-free shear acceleration tensor%
\index{shear@shear!torsion-free and curvature-free shear acceleration
tensor@torsion-free and curvature-free shear acceleration tensor} $_{sF}D_0$ 
\[
_{sF}D_0=\,_FD_0-\frac 1{n-1}\cdot \,_FU_0\cdot h_u 
\]
\begin{equation}
_FD_0=h_u(b_s)h_u%
\text{ ,\thinspace \thinspace \thinspace \thinspace \thinspace \thinspace
\thinspace \thinspace \thinspace \thinspace }_FU_0=g[b]-\frac 1e\cdot
g(u,\nabla _ua)\text{ .}  \label{9.10}
\end{equation}

In index form 
\begin{equation}
(_FD_0)_{ij}=(_FD_0)_{ji}=\frac 12\cdot h_{i\overline{k}}\cdot (a^k\text{ }%
_{;n}\cdot g^{nl}+a^l\text{ }_{;n}\cdot g^{nk})\cdot h_{\overline{l}j}\text{
,}  \label{9.11}
\end{equation}
\begin{equation}
\begin{array}{c}
_FU_0=a^k\text{ }_{;k}-\frac 1e\cdot g_{\overline{k}\overline{l}}\cdot
u^k\cdot a^l\text{ }_{;m}\cdot u^m= \\ 
=a^k\text{ }_{;k}-\frac 1e\cdot [(g_{\overline{k}\overline{l}}\cdot u^k\cdot
a^l)_{;m}\cdot u^m-g_{kl;m}\cdot u^m\cdot u^{\overline{k}}\cdot a^{\overline{%
l}}-g_{\overline{k}\overline{l}}\cdot a^k\cdot a^l]\text{ .}
\end{array}
\label{9.12}
\end{equation}

(f) Torsion-free and curvature-free rotation acceleration tensor%
\index{rotation@rotation!torsion-free and curvature-free rotation
acceleration@torsion-free and curvature-free rotation acceleration} $_FW_0$%
\begin{equation}
_FW_0=h_u(b_a)h_u%
\text{ .}  \label{9.13}
\end{equation}

In index form 
\begin{equation}
(_FW_0)_{ij}=-(_FW_0)_{ji}=\frac 12\cdot h_{i\overline{k}}\cdot (a^k\text{ }%
_{;n}\cdot g^{nl}-a^l\text{ }_{;n}\cdot g^{nk})\cdot h_{\overline{l}j}.
\label{9.14}
\end{equation}

(g) Torsion-free and curvature-free expansion acceleration%
\index{expansion@expansion!torsion-free and curvature-free
expansion@torsion-free and curvature-free expansion} $_FU_0$ [s. (e)].

(h) Curvature-free shear acceleration tensor%
\index{shear@shear!curvature-free shear acceleration tensor@curvature-free
shear acceleration tensor} $_sD_0=D_0-\frac 1{n-1}\cdot U_0\cdot h_u$%
\begin{equation}
\begin{array}{c}
D_0=h_u(b_s)h_u-\frac 12\cdot [_sP(%
\overline{g})\sigma +\sigma (\overline{g})_sP]-\frac 12\cdot [Q(\overline{g}%
)\omega +\omega (\overline{g})Q]- \\ 
-\frac 1{n-1}\cdot (\theta _1\cdot \sigma +\theta \cdot \,_sP)-\frac
1{n-1}\cdot (\theta _1^{.}+\frac 1{n-1}\cdot \theta _1\cdot \theta )\cdot
h_u-\nabla _u(_sP)- \\ 
-\frac 12\cdot [_sP(\overline{g})\omega -\omega (\overline{g})_sP]-\frac
12\cdot [Q(\overline{g})\sigma -\sigma (\overline{g})Q]- \\ 
-\frac 1{2\cdot e}\cdot [h_u(a)\otimes
(g(u))(m+q)h_u+h_u((g(u))(m+q))\otimes h_u(a)]- \\ 
-\frac 1e\cdot [_sP(a)\otimes g(u)+g(u)\otimes \,_sP(a)]- \\ 
-\frac 12\cdot [h_u(\nabla _u\overline{g})_sP+\,_sP(\nabla _u\overline{g}%
)h_u]-\frac 12\cdot [h_u(\nabla _u\overline{g})Q-Q(\nabla _u\overline{g})h_u]%
\text{ .}
\end{array}
\label{9.15}
\end{equation}

In index form 
\begin{equation}
\begin{array}{c}
(D_0)_{ij}=(D_0)_{ji}=h_{\overline{k}(i}\cdot h_{j)\overline{l}}\cdot a^k%
\text{ }_{;m}\cdot g^{ml}-\,_sP_{k(i}\cdot \sigma _{j)l}\cdot g^{\overline{k}%
\overline{l}}-Q_{k(i}\cdot \omega _{j)l}\cdot g^{\overline{k}\overline{l}}-
\\ 
-\frac 1{n-1}\cdot (\theta _1\cdot \sigma _{ij}+\theta \cdot
\,_sP_{ij})-\frac 1{n-1}\cdot (\theta _1^{.}+\frac 1{n-1}\cdot \theta
_1\cdot \theta )\cdot h_{ij}- \\ 
-\,_sP_{ij;m}\cdot u^m+\,_sP_{k(i}\cdot \omega _{j)l}\cdot g^{\overline{k}%
\overline{l}}+Q_{k(i}\cdot \sigma _{j)l}\cdot g^{\overline{k}\overline{l}}-
\\ 
-\frac 1e\cdot a^k\cdot [_sP_{i\overline{k}}\cdot u_j+\,_sP_{j\overline{k}%
}\cdot u_i+h_{\overline{k}(i}\cdot h_{j)\overline{l}}\cdot u_{\overline{n}%
}\cdot T_{mr}\,^n\cdot u^r\cdot g^{ml}]- \\ 
-\,_sP_{\overline{k}(i}\cdot h_{j)\overline{l}}\cdot g^{kl}\text{ }%
_{;m}\cdot u^m-Q_{\overline{k}(i}\cdot h_{j)\overline{l}}\cdot g^{kl}\text{ }%
_{;m}\cdot u^m\text{ .}
\end{array}
\label{9.16}
\end{equation}

(i) Curvature-free rotation acceleration tensor%
\index{rotation@rotation!curvature-free rotation acceleration
tensor@curvature-free rotation acceleration tensor} $W_0$%
\begin{equation}
\begin{array}{c}
W_0=h_u(b_a)h_u-\frac 12\cdot [_sP(%
\overline{g})\sigma -\sigma (\overline{g})_sP]-\frac 12\cdot [Q(\overline{g}%
)\omega -\omega (\overline{g})Q]- \\ 
-\frac 1{n-1}\cdot (\theta _1\cdot \omega +\theta \cdot Q)-\nabla _uQ-\frac
12\cdot [_sP(\overline{g})\omega +\omega (\overline{g})_sP]- \\ 
-\frac 12\cdot [Q(\overline{g})\sigma +\sigma (\overline{g})Q]-\frac 1e\cdot
[Q(a)\otimes g(u)-g(u)\otimes Q(a)]- \\ 
-\frac 1{2\cdot e}\cdot [h_u(a)\otimes
(g(u))(m+q)h_u-h_u((g(u))(m+q))\otimes h_u(a)]- \\ 
-\frac 12\cdot [h_u(\nabla _u\overline{g})_sP-\,_sP(\nabla _u\overline{g}%
)h_u]-\frac 12\cdot [h_u(\nabla _u\overline{g})Q+Q(\nabla _u\overline{g})h_u]%
\text{ .}
\end{array}
\label{9.17}
\end{equation}

In index form 
\begin{equation}
\begin{array}{c}
(W_0)_{ij}=-(W_0)_{ji}=h_{\overline{k}[i}\cdot h_{j]\overline{l}}\cdot a^k%
\text{ }_{;m}\cdot g^{ml}-\,_sP_{\overline{k}[i}\cdot \sigma _{j]\overline{l}%
}\cdot g^{kl}-Q_{\overline{k}[i}\cdot \omega _{j]\overline{l}}\cdot g^{kl}-
\\ 
-\frac 1{n-1}\cdot (\theta _1\cdot \omega _{ij}+\theta \cdot
Q_{ij})-Q_{ij;m}\cdot u^m+ \\ 
+_sP_{k[i}\cdot \omega _{j]l}\cdot g^{\overline{k}\overline{l}}+Q_{k[i}\cdot
\sigma _{j]l}\cdot g^{\overline{k}\overline{l}}- \\ 
-\frac 1e\cdot a^k\cdot (Q_{i\overline{k}}\cdot u_j-Q_{j\overline{k}}\cdot
u_i+h_{\overline{k}[i}\cdot h_{j]\overline{l}}\cdot u_{\overline{n}}\cdot
T_{mr}\,^n\cdot u^r\cdot g^{ml})+ \\ 
+\,_sP_{\overline{k}[i}\cdot h_{j]\overline{l}}\cdot g^{kl}\text{ }%
_{;m}\cdot u^m+Q_{\overline{k}[i}\cdot h_{j]\overline{l}}\cdot g^{kl}\text{ }%
_{;m}\cdot u^m\text{ .}
\end{array}
\label{9.18}
\end{equation}

(j) Curvature-free expansion acceleration%
\index{expansion@expansion!curvature-free expansion
acceleration@curvature-free expansion acceleration} $U_0$%
\begin{equation}
\begin{array}{c}
U_0=g[b]-%
\overline{g}[_sP(\overline{g})\sigma ]-\overline{g}[Q(\overline{g})\omega %
]-\theta _1^{.}-\frac 1{n-1}\cdot \theta _1\cdot \theta - \\ 
-\frac 1e\cdot [g(u,T(a,u))+g(u,\nabla _ua)]\text{ .}
\end{array}
\label{9.19}
\end{equation}

In index form 
\begin{equation}
\begin{array}{c}
U_0=a^k\text{ }_{;k}-g^{\overline{i}\overline{j}}\text{ }\cdot
\,_sP_{ik}\cdot g^{\overline{k}\overline{l}}\cdot \sigma _{lj}-g^{\overline{i%
}\overline{j}}\cdot Q_{ik}\cdot g^{\overline{k}\overline{l}}\cdot \omega
_{lj}-\theta _1^{.}-\frac 1{n-1}\cdot \theta _1\cdot \theta - \\ 
-\frac 1e\cdot [a^k\cdot (u_{\overline{n}}\cdot T_{km}\,^n\cdot u^m-2\cdot
g_{\overline{k}\overline{m};l}\cdot u^l\cdot u^m-g_{\overline{k}\overline{l}%
}\cdot a^l)+ \\ 
+\frac 12\cdot (e_{,k}\cdot u^k)_{,l}\cdot u^l-\frac 12\cdot (g_{mn;r}\cdot
u^r)_{;s}\cdot u^s\cdot u^{\overline{m}}\cdot u^{\overline{n}}]\text{ .}
\end{array}
\label{9.20}
\end{equation}

(k) Shear acceleration tensor induced by the torsion%
\index{shear@shear!shear acceleration induced by the torsion@shear
acceleration induced by the torsion} $_{sT}D_0$%
\[
_{sT}D_0=\,_TD_0-\frac 1{n-1}\cdot \,_TU_0\cdot h_u 
\]

\begin{equation}
\begin{array}{c}
_TD_0=\frac 12\cdot [_sP(%
\overline{g})\sigma +\sigma (\overline{g})_sP]+\frac 12\cdot [Q(\overline{g}%
)\omega +\omega (\overline{g})Q]+ \\ 
+\frac 1{n-1}\cdot (\theta _1\cdot \sigma +\theta \cdot \,_sP)+\frac
1{n-1}\cdot (\theta _1^{.}+\frac 1{n-1}\cdot \theta _1\cdot \theta )\cdot
h_u+\nabla _u(_sP)+ \\ 
+\frac 12\cdot [_sP(\overline{g})\omega -\omega (\overline{g})_sP]+\frac
12\cdot [Q(\overline{g})\sigma -\sigma (\overline{g})Q]+ \\ 
+\frac 1{2\cdot e}\cdot [h_u(a)\otimes
(g(u))(m+q)h_u+h_u((g(u))(m+q))\otimes h_u(a)]+ \\ 
+\frac 1e\cdot [_sP(a)\otimes g(u)+g(u)\otimes _sP(a)]+ \\ 
+\frac 12\cdot [h_u(\nabla _u\overline{g})_sP+\,_sP(\nabla _u\overline{g}%
)h_u]+\frac 12\cdot [h_u(\nabla _ug)Q-Q(\nabla _ug)h_u]\text{ .}
\end{array}
\label{9.21}
\end{equation}

In index form 
\begin{equation}
(_TD_0)_{ij}=(_FD_0)_{ij}-(D_0)_{ij}\text{ .}  \label{9.22}
\end{equation}

(l) Expansion acceleration induced by the torsion%
\index{expansion@expansion!expansion acceleration induced by the
torsion@expansion acceleration induced by the torsion} $_TU_0$%
\begin{equation}
_TU_0=%
\overline{g}[_sP(\overline{g})\sigma ]+\overline{g}[Q(\overline{g})\omega %
]+\theta _1^{.}+\frac 1{n-1}\cdot \theta _1\cdot \theta +\frac 1e\cdot
g(u,T(a,u))\text{ .}  \label{9.23}
\end{equation}

In index form 
\[
_TU_0=\,_FU_0-U_0\text{ .} 
\]

(m) Rotation acceleration tensor induced by the torsion%
\index{rotation@rotation!rotation acceleration induced by the
torsion@rotation acceleration induced by the torsion} $_TW_0$%
\begin{equation}
\begin{array}{c}
_TW_0=\frac 12\cdot [_sP(%
\overline{g})\sigma -\sigma (\overline{g})_sP]+\frac 12\cdot [Q(\overline{g}%
)\omega -\omega (\overline{g})Q)]+ \\ 
+\frac 1{n-1}\cdot (\theta _1\cdot \omega +\theta \cdot Q)+\nabla _uQ+\frac
12\cdot [_sP(\overline{g})\omega +\omega (\overline{g})_sP]+ \\ 
+\frac 12\cdot [Q(\overline{g})\sigma +\sigma (\overline{g})Q]+ \\ 
+\frac 1{2\cdot e}\cdot [h_u(a)\otimes
(g(u))(m+q)h_u-h_u((g(u))(m+q))\otimes h_u(a)]+ \\ 
+\frac 1e\cdot [Q(a)\otimes g(u)-g(u)\otimes Q(a)]+ \\ 
+\frac 12\cdot [h_u(\nabla _u\overline{g})_sP-\,_sP(\nabla _u\overline{g}%
)h_u]+\frac 12\cdot [h_u(\nabla _u\overline{g})Q+Q(\nabla _u\overline{g})h_u]%
\text{ .}
\end{array}
\label{9.24}
\end{equation}

In index form 
\begin{equation}
(_TW_0)_{ij}=(_FW_0)_{ij}-(W_0)_{ij}\text{ .}  \label{9.25}
\end{equation}

(n) Shear acceleration tensor induced by the curvature%
\index{shear@shear!shear acceleration induced by the curvature@shear
acceleration induced by the curvature} $_sM=M-\frac 1{n-1}\cdot I\cdot h_u$ 
\begin{equation}
\begin{array}{c}
M=\frac 1e\cdot h_u(a)\otimes h_u(a)+\frac 12\cdot [_sE(%
\overline{g})\sigma +\sigma (\overline{g})_sE]+\frac 12\cdot [S(\overline{g}%
)\omega +\omega (\overline{g})S]+ \\ 
+\frac 1{n-1}\cdot (\theta _o\cdot \sigma +\theta \cdot \,_sE)+\frac
1{n-1}\cdot (\theta _o^{.}+\frac 1{n-1}\cdot \theta _o\cdot \theta
)h_u+\nabla _u(_sE)+ \\ 
+\frac 12\cdot [_sE(\overline{g})\omega -\omega (\overline{g})_sE]+\frac
12\cdot [S(\overline{g})\sigma -\sigma (\overline{g})S]+ \\ 
+\frac 1{2\cdot e}\cdot [h_u(a)\otimes (g(u))(k_0+k-\nabla _u\overline{g}%
)h_u+h_u((g(u))(k_0+k-\nabla _u\overline{g}))\otimes h_u(a)]+ \\ 
+\frac 1e\cdot [_sE(a)\otimes g(u)+g(u)\otimes \,_sE(a)]+ \\ 
+\frac 12\cdot [h_u(\nabla _u\overline{g})_sE+\,_sE(\nabla _u\overline{g}%
)h_u]+\frac 12\cdot [h_u(\nabla _u\overline{g})S-S(\nabla _u\overline{g}%
)h_u]- \\ 
-h_u(b_s)h_u\text{ .}
\end{array}
\label{9.26}
\end{equation}

In index form 
\begin{equation}
\begin{array}{c}
M_{ij}=M_{ji}=\frac 1e\cdot h_{i\overline{k}}\cdot a^k\cdot a^l\cdot h_{%
\overline{l}j}+\,_sE_{k(i}\cdot \sigma _{j)l}\cdot g^{\overline{k}\overline{l%
}}+S_{k(i}\cdot \omega _{j)l}\cdot g^{\overline{k}\overline{l}}+ \\ 
+\frac 1{n-1}\cdot (\theta _o\cdot \sigma _{ij}+\theta \cdot
\,_sE_{ij})+\frac 1{n-1}\cdot (\theta _o^{.}+\frac 1{n-1}\cdot \theta
_o\cdot \theta )\cdot h_{ij}- \\ 
-\,_sE_{k(i}\cdot \omega _{j)l}\cdot g^{\overline{k}\overline{l}%
}-S_{k(i}\cdot \sigma _{j)l}\cdot g^{\overline{k}\overline{l}%
}+\,_sE_{ij;k}\cdot u^k+ \\ 
+\frac 1e\cdot a^k\cdot [_sE_{i\overline{k}}\cdot u_j+\,_sE_{j\overline{k}%
}\cdot u_i+h_{\overline{k}(i}\cdot h_{j)\overline{l}}\cdot g^{ml}\cdot
(e_{,m}-u_{\overline{n}}\cdot T_{mr}\,^n\cdot u^r- \\ 
-g_{rs;m}\cdot u^{\overline{r}}\cdot u^{\overline{s}}+g_{\overline{m}%
\overline{r};s}\cdot u^s\cdot u^r)]+ \\ 
+\,_sE_{\overline{k}(i}\cdot h_{j)\overline{l}}\cdot g^{kl}\text{ }%
_{;s}\cdot u^s+S_{\overline{k}(i}\cdot h_{j)\overline{l}}\cdot g^{kl}\text{ }%
_{;s}\cdot u^s- \\ 
-h_{\overline{k}(i}\cdot h_{j)\overline{l}}\cdot a^k\text{ }_{;m}\cdot g^{ml}%
\text{ .}
\end{array}
\label{9.27}
\end{equation}

(o) Expansion acceleration induced by the curvature%
\index{expansion@expansion!expansion acceleration induced by the
curvature@expansion acceleration induced by the curvature} $I$%
\begin{equation}
\begin{array}{c}
I=-g[b]+%
\overline{g}[_sE(\overline{g})\sigma ]+\overline{g}[S(\overline{g})\omega %
]+\theta _o^{.}+\frac 1{n-1}\cdot \theta _o\cdot \theta + \\ 
+\frac 1e\cdot [2\cdot g(u,\nabla _au)-g(u,T(a,u))+u(g(u,a))]- \\ 
-\frac 1{e^2}\cdot g(u,a)\cdot [3\cdot g(u,a)+(\nabla _ug)(u,u)]\text{ .}
\end{array}
\label{9.28}
\end{equation}

In index form 
\begin{equation}
\begin{array}{c}
I=R_{ij}\cdot u^i\cdot u^j=-a^j\text{ }_{;j}+g^{\overline{i}\overline{j}%
}\cdot g^{\overline{k}\overline{l}}\cdot \,_sE_{ik}\cdot \sigma _{lj}+g^{%
\overline{i}\overline{j}}\cdot g^{\overline{k}\overline{l}}\cdot S_{ik}\cdot
\omega _{lj}+ \\ 
+\theta _o^{.}+\frac 1{n-1}\cdot \theta _o\cdot \theta +\frac 1e\cdot
[a^k\cdot (e_{,k}-u_{\overline{n}}\cdot T_{km}\,^n\cdot u^m-g_{mn;k}\cdot u^{%
\overline{m}}\cdot u^{\overline{n}}- \\ 
-g_{\overline{k}\overline{m};s}\cdot u^s\cdot u^m)+\frac 12\cdot (u^k\cdot
e_{,k})_{,l}\cdot u^l-\frac 12\cdot (g_{mn;r}\cdot u^r)_{;s}\cdot u^s\cdot
u^{\overline{m}}\cdot u^{\overline{n}}]- \\ 
-\frac 1{e^2}\cdot [\frac 34\cdot (e_{,k}\cdot u^k)^2-(e_{,k}\cdot u^k)\cdot
g_{mn;r}\cdot u^r\cdot u^{\overline{m}}\cdot u^{\overline{n}}+\frac 14\cdot
(g_{mn;r}\cdot u^r\cdot u^{\overline{m}}\cdot u^{\overline{n}})^2]\text{ .}
\end{array}
\label{9.29}
\end{equation}

(p) Rotation acceleration tensor induced by the curvature%
\index{rotation@rotation!rotation acceleration induced by the
curvature@rotation acceleration induced by the curvature} $N$%
\begin{equation}
\begin{array}{c}
N=\frac 12\cdot [_sE(%
\overline{g})\sigma -\sigma (\overline{g})_sE]+\frac 12\cdot [S(\overline{g}%
)\omega -\omega (\overline{g})S]+ \\ 
+\frac 1{n-1}\cdot (\theta _o\cdot \omega +\theta \cdot S)+\nabla _uS+\frac
12\cdot [_sE(\overline{g})\omega +\omega (\overline{g})_sE]+ \\ 
+\frac 12\cdot [S(\overline{g})\sigma +\sigma (\overline{g})S]+ \\ 
+\frac 1{2\cdot e}\cdot [h_u(a)\otimes (g(u))(k_0+k-\nabla _u\overline{g}%
)h_u-h_u((g(u))(k_0+k-\nabla _u\overline{g}))\otimes h_u(a)]+ \\ 
+\frac 1e\cdot [S(a)\otimes g(u)-g(u)\otimes S(a)]+ \\ 
+\frac 12\cdot [h_u(\nabla _u\overline{g})_sE-\,_sE(\nabla _u\overline{g}%
)h_u]+\frac 12\cdot [h_u(\nabla _u\overline{g})S+S(\nabla _u\overline{g}%
)h_u]- \\ 
-h_u(b_a)h_u\text{ .}
\end{array}
\label{9.30}
\end{equation}

In index form 
\begin{equation}
\begin{array}{c}
N_{ij}=-N_{ji}=\,_sE_{k[i}\cdot \sigma _{j]l}\cdot g^{\overline{k}\overline{l%
}}+S_{k[i}\cdot \omega _{j]l}\cdot g^{\overline{k}\overline{l}}+\frac
1{n-1}\cdot (\theta _o\cdot \omega _{ij}+\theta \cdot S_{ij})- \\ 
-\,_sE_{k[i}\cdot \omega _{j]l}\cdot g^{\overline{k}\overline{l}%
}-S_{k[i}\cdot \sigma _{j]l}\cdot g^{\overline{k}\overline{l}}+S_{ij;k}\cdot
u^k-h_{\overline{k}[i}\cdot h_{j]\overline{l}}\cdot a^k\text{ }_{;m}\cdot
g^{ml}+ \\ 
+\frac 1e\cdot a^k\cdot [S_{i\overline{k}}\cdot u_j-S_{j\overline{k}}\cdot
u_i+h_{\overline{k}[i}\cdot h_{j]\overline{l}}\cdot g^{ml}\cdot (e_{,m}-u_{%
\overline{n}}\cdot T_{mr}\,^n\cdot u^r- \\ 
-g_{rs;m}\cdot u^{\overline{r}}\cdot u^{\overline{s}}+g_{\overline{m}%
\overline{r};s}\cdot u^s\cdot u^r)]-\,_sE_{\overline{k}[i}\cdot h_{j]%
\overline{l}}\cdot g^{kl}\text{ }_{;s}\cdot u^s-S_{\overline{k}[i}\cdot h_{j]%
\overline{l}}\cdot g^{kl}\text{ }_{;s}\cdot u^s\text{ .}
\end{array}
\label{9.31}
\end{equation}

\section{Table 1. Kinematic characteristics connected with the notions
relative velocity and relative acceleration. A summary of the definitions}

\subsection{Kinematic characteristics connected with the relative velocity}

\medskip

1. Relative position vector field%
\index{vector field@vector field!relative position vector field@relative
position vector field}

(relative position vector) ............................................... $%
\xi _{\perp }=%
\overline{g}(h_u(\xi ))$

2. Relative velocity ................................................. $%
_{rel}v=\overline{g}(h_u(\nabla _u\xi ))$

3. Deformation velocity tensor

(deformation velocity, deformation) .. $d=d_0-d_1=\sigma +\omega +\frac
1{n-1}\cdot \theta \cdot h_u$

4. Torsion-free deformation velocity tensor

(torsion-free deformation velocity, torsion-free deformation)

................................................................. $%
d_0=\,_sE+S+\frac 1{n-1}\cdot \theta _o\cdot h_u$

5. Deformation velocity tensor induced by the torsion

(torsion deformation velocity, torsion deformation)

................................................................. $%
d_1=\,_sP+Q+\frac 1{n-1}\cdot \theta _1\cdot h_u$

6. Shear velocity tensor

(shear velocity, shear)
....................................................... $\sigma =\,_sE-\,_sP$

7. Torsion-free shear velocity tensor

(torsion shear velocity, torsion shear) ................. $_sE=E-\frac
1{n-1}\cdot \theta _o\cdot h_u$

8. Shear velocity tensor induced by the torsion

(torsion shear velocity tensor, torsion shear velocity, torsion shear)

.......................................................................... $%
_sP=P-\frac 1{n-1}\cdot \theta _1\cdot h_u$

9. Rotation velocity tensor

(rotation velocity, rotation)
................................................... $\omega =S-Q$

10. Torsion-free rotation velocity tensor

(torsion-free rotation velocity, torsion-free rotation)

.......................................................................................... 
$S=h_u(s)h_u$

11. Rotation velocity tensor induced by the torsion

(torsion rotation velocity, torsion rotation) ......................... $%
Q=h_u(q)h_u$

12. Expansion velocity

(expansion)
....................................................................... $%
\theta =\theta _o-\theta _1$

13. Torsion-free expansion velocity

(torsion-free expansion)
...................................................... $\theta _o=\overline{g%
}[E]$

14. Expansion velocity induced by the torsion

(torsion expansion velocity, torsion expansion) ..................... $%
\theta _1=\overline{g}[P]$

\subsection{Kinematic characteristics connected with the relative
acceleration}

\medskip

1. Acceleration%
\index{acceleration@acceleration}
......................................................................... $%
a=\nabla _uu$

2. Relative acceleration ................... ...................... $%
_{rel}a= $ $%
\overline{g}(h_u(\nabla _u\nabla _u\xi ))$

3. Deformation acceleration tensor

(deformation acceleration) ............................ $A=\,_sD+W+\frac
1{n-1}\cdot U\cdot h_u$

........................................................................................... 
$A=A_0+G$

.............................................................................. 
$A=\,_FA_0-\,_TA_0+G$

4. Torsion-free and curvature-free deformation acceleration tensor

(torsion-free and curvature-free deformation acceleration)

..................................................... $_FA_0=\,_{sF}D_0+%
\,_FW_0+\frac 1{n-1}\cdot \,_FU_0\cdot h_u$

4.a. Curvature-free deformation acceleration tensor

(curvature-free deformation acceleration). $A_0=\,_sD_0+W_0+\frac
1{n-1}\cdot U_0\cdot h_u$

5. Deformation acceleration tensor induced by the torsion

(torsion deformation acceleration tensor, torsion deformation acceleration)

..................................................... $\,\,_TA_0=\,_{sT}D_0+%
\,_TW_0+\frac 1{n-1}\cdot \,_TU_0\cdot h_u$

5.a. Deformation acceleration tensor induced by the curvature

(curvature deformation acceleration tensor, curvature deformation
acceleration)

...................................................................... $%
G=\,_sM+N+\frac 1{n-1}\cdot I\cdot h_u$

6. Shear acceleration tensor

(shear acceleration) ............................................. $%
_sD=D-\frac 1{n-1}\cdot U\cdot h_u$

.................................................................................... 
$_sD=\,_sD_0+_sM$

...................................................................... $%
_sD=\,_{sF}D_0-\,_{sT}D_0+\,_sM$

7. Torsion-free and curvature-free shear acceleration tensor

(torsion-free and curvature-free shear acceleration)

............................................................... $%
_{sF}D_0=\,_FD_0-\frac 1{n-1}\cdot \,_FU_0\cdot h_u$

7.a. Curvature-free shear acceleration tensor

(curvature-free shear acceleration) ................ $_sD_0=D_0-\frac
1{n-1}\cdot U_0\cdot h_u$

.............................................................................. 
$_sD_0=\,_{sF}D_0-\,_{sT}D_0$

8. Shear acceleration tensor induced by the torsion

(torsion shear acceleration tensor, torsion shear acceleration)

............................................................... $%
_{sT}D_0=\,_T\,D_0-\frac 1{n-1}\cdot \,_TU_0\cdot h_u$

8.a. Shear acceleration tensor induced by the curvature

(curvature shear acceleration tensor, curvature shear acceleration)

........................................................................... $%
_sM=M-\frac 1{n-1}\cdot I\cdot h_u$

9. Rotation acceleration tensor

(rotation acceleration)
..................................................... $W=W_0+N$

......................................................................... $%
W=\,_FW_0-\,_TW_0+N$

10. Torsion-free and curvature-free rotation acceleration tensor

(torsion-free and curvature-free rotation acceleration)

................................................................................. 
$_FW_0=h_u(b_a)h_u$

10.a. Curvature-free rotation acceleration tensor

(curvature-free rotation acceleration) .............................. $%
W_0=W-N$

.............................................................................. 
$W_0=\,_FW_0-\,_TW_0$

11. Rotation acceleration tensor induced by the torsion

(torsion rotation acceleration tensor, torsion rotation acceleration)

.............................................................................. 
$_TW_0=\,_FW_0-W_0$

11.a. Rotation acceleration tensor induced by the curvature

(curvature rotation acceleration tensor, curvature rotation acceleration)

.................................................................................... 
$N=h_u(K_a)h_u$

12. Expansion acceleration .............................................. $%
U=U_0+I$

.......................................................................... $%
U=\,_FU_0-\,_TU_0+I$

13. Torsion-free and curvature-free expansion acceleration

.................................................................................. 
$_FU_0=\overline{g}[_FD_0]$

13.a. Curvature-free expansion acceleration .................... $U_0=%
\overline{g}[D_0]$

.............................................................................. 
$\,\,U_0=\,_FU_0-\,_TU_0$

14. Expansion acceleration induced by the torsion

(torsion expansion acceleration) .................................... $U_0=%
\overline{g}[_TD_0]$

14.a. Expansion acceleration induced by the curvature

(curvature expansion acceleration) ........................... $I=\overline{g%
}[M]=\overline{g}[G]$

\section{Table 2. Classification of non-isotropic auto-parallel vector
fields on the basis of the kinematic characteristics connected with the
relative velocity and relative acceleration}

\subsection{Classification on the basis of kinematic characteristics
connected with the relative velocity}

The following conditions, connected with the relative velocity, can
characterize the vector fields over manifolds with affine connections and
metrics:

1. $\sigma =0$.

2. $\omega =0$.

3. $\theta =0$.

4. $\sigma =0$, $\omega =0$.

5. $\sigma =0$, $\theta =0$.

6. $\omega =0$, $\theta =0$.

7. $\sigma =0$, $\omega =0$, $\theta =0$.

8. $_sE=0$.

9. $S=0$.

10. $\theta _o=0$.

11. $_sE=0$, $S=0$.

12. $_sE=0$, $\theta _o=0$.

13. $S=0$, $\theta _o=0$.

14. $_sE=0$, $S=0$, $\theta _o=0$.

15. $_sP=0.$

16. $Q=0$.

17. $\theta _1=0$.

18. $_sP=0$, $Q=0$.

19. $_sP=0$, $\theta _1=0$.

20. $Q=0$, $\theta _1=0$.

21. $_sP=0$, $Q=0$, $\theta _1=0$.

\subsection{Classification on the basis of kinematic characteristics
connected with the relative acceleration}

The following conditions, connected with the relative acceleration, can
characterize the vector fields over manifolds with affine connections and
metrics:

1. $_sD=0$.

2. $W=0$.

3. $U=0$.

4. $_sD=0$, $W=0$.

5. $_sD=0$, $U=0$.

6. $W=0$, $U=0$.

7. $_sD=0$, $W=0$, $U=0$.

8. $_sM=0$.

9. $N=0$.

10. $I=0$.

11. $_sM=0$, $N=0$.

12. $_sM=0$, $I=0$.

13. $N=0$, $I=0$.

14. $_sM=0$, $N=0$, $I=0$.

15. $_{sT}D_0=0$.

16. $_TW_0=0$.

17. $_TU_0=0$.

18. $_{sT}D_0=0$, $_TW_0=0$.

19. $_{sT}D_0=0$, $_TU_0=0$.

20. $_TW_0=0$, $_TU_0=0$.

21. $_{sT}D_0=0$, $_TW_0=0$, $_TU_0=0$.

The kinematic characteristics related to the relative velocity and the
friction velocity are related to deformations and tensions. The last
statement will be discussed in the next paper.

\section{Conclusion}

In the present paper the notion of relative acceleration is introduced and
its corresponding kinematic characteristics (deformation acceleration, shear
acceleration, rotation acceleration,expansion acceleration) are considered.
On their basis classification of contravariant vector fields are proposed
for describing flows with special properties and motions. The consideration
are important for developing appropriate models of flows and continuous
media in relativistic and more general (gravitational) field theories.

\end{document}